\documentclass[aps,prc,twocolumn,floatfix,nofootinbib]{revtex4-2}

\usepackage{graphicx}
\usepackage[urlcolor=green]{hyperref}
\usepackage{amsmath}
\usepackage{bm}

\usepackage{color}

\newcommand{\Nresc}{N_{\rm resc.}}

\begin{document}

\author{Benedikt Bachmann}
\author{Nicolas Borghini} \email{borghini@physik.uni-bielefeld.de}
\author{Nina Feld} \email{nkersting@uni-bielefeld.de}
\author{Hendrik Roch} \email{hroch@physik.uni-bielefeld.de}
\affiliation{Fakult\"at f\"ur Physik, Universit\"at Bielefeld, 
D-33615 Bielefeld, Germany}

\title{On differences between even and odd anisotropic-flow harmonics in non-equilibrated systems}

\begin{abstract}
To assess how anisotropic transverse flow is created in a system out of equilibrium, we compare several kinetic-theoretical models in the few-rescatterings regime.
We compare the flow harmonics $v_n$ from three types of transport simulations, with either $2\to 2$ or $2\to 0$ collision kernels and in the former case allowing the particles to rescatter several times or not, and from analytical calculations neglecting the gain term of the Boltzmann equation. 
We find that the even flow harmonics are similar in all approaches, while the odd ones differ significantly. 
This suggests that while even $v_n$ harmonics may to a large extent be due to the anisotropic escape probability of particles, this is not the predominant mechanism underlying the odd $v_n$ coefficients. 
\end{abstract}

\maketitle

\section{Introduction}
\label{s:intro}

The charged hadrons produced in collisions of heavy nuclei at high energy show a characteristic azimuthally asymmetric transverse emission pattern~\cite{Heinz:2013th}. 
This aniso\-tropic flow, usually quantified in terms of coefficients in the Fourier expansion of the transverse momentum distributions~\cite{Voloshin:1994mz}, has also been observed in so-called smaller systems, namely proton-- and deuteron--nucleus or even proton--proton collisions with large multiplicities~\cite{Nagle:2018nvi}.

The anisotropic flow results have been essential for singling out relativistic hydrodynamics~\cite{Jeon:2015dfa,Romatschke:2017ejr} as the model of choice for describing the dynamics of the system created in heavy-ion collisions, which is then modeled as a continuous medium, whose initial geometrical asymmetry is converted in the evolution into the final state momentum anisotropy~\cite{Ollitrault:1992bk}.
Indeed, relativistic fluid dynamics can describe satisfactorily a large amount of data on anisotropic flow and related azimuthal correlations~\cite{Huovinen:2006jp,Teaney:2009qa,Heinz:2013th,Luzum:2013yya}.

However, the question of the applicability of fluid dynamics is still under discussion, in particular when the number of emitted particles is small~\cite{Weller:2017tsr,Zhao:2020pty}.
Thus, alternative descriptions based on microscopic kinetic transport theory, which is known to reproduce fluid-dynamical results when particles undergo many rescatterings~\cite{Gombeaud:2007ub}, are being explored again, in particular with a view to small systems. 
A number of recent attempts start from semi-realistic initial geometries, which allow to isolate the flow harmonics of interest and study their origin~\cite{Romatschke:2018wgi,Borghini:2018xum,Kurkela:2018qeb,Kurkela:2019kip,Kurkela:2020wwb,Kurkela:2021ctp,Ambrus:2021fej,Borghini:2022qha}.

In one of the more realistic transport studies~\cite{He:2015hfa}, within the AMPT approach, it was claimed that ``the majority'' of the measured anisotropic flow signal (for elliptic flow $v_2$ and triangular flow $v_3$ in Au--Au collisions at RHIC energy) is not due to the numerous rescatterings of the produced particles, but could rather be dominated by those particles that escape the asymmetric system geometry without having scattered. 
Primitive versions of this ``escape mechanism'' scenario had been considered earlier with simple initial states allowing analytical calculations with simplifying assumptions~\cite{Heiselberg:1998es,Borghini:2010hy}, and also used for an early estimate of the $v_2$ of $J/\psi$ quarkonia~\cite{Wang:2002ck}.%

Yet recent findings cast some doubt on the efficiency of the anisotropic-escape picture in the regime of very few rescatterings, especially regarding $v_3$. 
Thus, it was found in Ref.~\cite{Kurkela:2021ctp} that the value of $v_3$ --- to be accurate, of energy-weighted triangular flow --- in kinetic models at low opacity, i.e.\ when particles rescatter very little, depends significantly on the collision kernel of the Boltzmann equation: 
triangular flow (divided by the initial triangularity) comes out negative in an effective kinetic theory of QCD, while it is positive in the relaxation-time approximation. 
In contrast, the behavior of elliptic flow seems to be more robust across scenarios. 

In this paper, we want to further explore the production of anisotropic flow in the regime of very few rescatterings, in particular with a view to testing the anisotropic-escape scenario. 
For that purpose, we employ numerical transport simulations with various collision kernels, in particular with elastic binary collisions (Sect.~\ref{s:methods}), complemented with analytical calculations that only account for the loss term of the Boltzmann equation. 
We then compare in Sect.~\ref{s:Results} the results for the $v_n$ coefficients in our various approaches and with those of the recent literature, before concluding in Sect.~\ref{s:summary}.

Since we focus on systems with very few rescatterings, the flow coefficients are at times very small. 
Accordingly, their values in transport simulations are likely to be affected by numerical fluctuations. 
High statistics are needed to counteract this noise, which is why we restrict ourselves to a two-dimensional system, to keep the computing time in reasonable bounds. 
This restriction will be further examined in Sect.~\ref{s:summary}.

\section{Methods}
\label{s:methods}

In order to investigate the importance of the ``escape mechanism'' for the production of anisotropic flow when particles undergo very few rescatterings, we perform four types of calculations.
On the one hand, numerical simulations with a transport code, with two different collision kernels: first a $2\to 2$ kernel that implements elastic binary collisions, and gives as reference the ``total'' anisotropic flow produced in a semi-realistic system. 
Secondly, a ``single-hit'' version using the $2\to 2$ kernel but in which particles that have already scattered once are no longer allowed to interact.
Thirdly, a $2\to 0$ collision kernel, such that the resulting flow is that of the particles that escaped the system without scattering. 
On the other hand, we perform analytical calculations within kinetic theory, using  only the loss term of the collision kernel in the Boltzmann equation, and working at linear order in the cross section: this provides a controlled approximation to the $2\to 0$ scenario, which itself includes all orders in the cross section.

We begin with introducing the analytical approach (Sect.~\ref{ss:analytical}), together with the initial conditions we use for both analytical and numerical calculations. 
We then briefly present our transport setups (Sect.~\ref{ss:numerics}). 
All calculations are performed with massless identical (yet distinguishable) particles, which propagate in two dimensions only, corresponding to the transverse plane in a high-energy nuclear collision. 
Two-dimensional vectors are denoted in boldface. 
Throughout the paper we use the convention $\hbar = c = 1$, and $(r,\theta)$ denote polar coordinates in the transverse plane, with their origin at the center of the system in its initial state.

\subsection{Analytical approach}
\label{ss:analytical}

In our analytical calculations, we characterize the particle system by a classical on-shell phase space distribution $f$, which obeys the relativistic Boltzmann equation%
\begin{equation}
\label{Boltzmann-eq}
p^\mu\partial_\mu f(t,{\bf x},{\bf p}) = {\cal C}[f(t,{\bf x},{\bf p})].
\end{equation}
Instead of a full collision kernel with detailed balance, we shall only consider the loss term of binary scatterings
\begin{equation}
\label{loss-term}
{\cal C}_{\rm loss}[f(t,{\bf x},{\bf p})] = -\frac{E_{\bf p}}{2}\!\int\!\!
  f(t,{\bf x},{\bf p})_{}f(t,{\bf x},{\bf p}_1)_{}v_{\rm rel.}\sigma\,{\rm d}^2{\bf p}_1,
\end{equation}
with $E_{\bf p}$ the energy of the particle with momentum ${\bf p}$,  $v_{\rm rel.}$ the M{\o}ller velocity, and $\sigma$ the total cross section.
Note that this implies that energy, momentum and particle number are not conserved in the evolution. 
For massless particles in two dimensions, $v_{\rm rel.} = 1-\cos(\varphi_{\bf p}-\varphi_1)$ where $\varphi_{\bf p}$ resp.\ $\varphi_1$ is the azimuthal angle of momentum ${\bf p}$ resp.\ ${\bf p}_1$.

The ``observables'' we study are the Fourier coefficients quantifying anisotropic flow~\cite{Voloshin:1994mz}, in particular their time evolution. 
In terms of the phase space distribution $f$, the momentum-integrated coefficients are given by%
\begin{equation}
\label{v_n(t)}
v_n(t) = \frac{\displaystyle\int\! f(t,{\bf x},{\bf p})  \cos(n \varphi_{\bf p}) \,{\rm d}^2{\bf x}\,{\rm d}^2{\bf p}}%
  {\displaystyle\int\! f(t,{\bf x},{\bf p})\,{\rm d}^2{\bf x}\,{\rm d}^2{\bf p}},
\end{equation}
where the denominator is simply the total number of particles $N(t)$ at time $t$. 
Differentiating this expression with respect to time gives two contributions, from the derivatives of the numerator and denominator respectively: 
\begin{align}
\partial_t v_n(t) = &\ \frac{1}{N(t)}\!\int\!\partial_t f(t,{\bf x},{\bf p})\cos(n \varphi_{\bf p})\,{\rm d}^2{\bf x}\,{\rm d}^2{\bf p} \cr
& - \frac{\partial_t N(t)}{N(t)}v_n(t).
\label{d_tv_n}
\end{align}
Using the Boltzmann equation to replace $\partial_t f$ in the integrand, the term involving the spatial gradient of $f$ gives zero after integrating over ${\bf x}$, since $f$ vanishes at infinity. 
There remains only the contribution from the collision term, which at leading order is a priori linear in $\sigma$. 
If we restrict ourselves to this linear order, as we do from now on, then we may neglect the change in $N(t)$ induced by the (particle-number non-conserving) rescatterings in the denominator in the first line of Eq.~\eqref{d_tv_n}, i.e.\ approximate $N(t)\simeq N(0)$, which we shall more briefly denote by $N$. 
In addition, we may also neglect the evolution of the phase-space density induced by rescatterings in the integrand of the collision term in the numerator. 
That is, we replace $f(t,{\bf x},{\bf p})$ by the free-streaming distribution $f_{\rm f.s.}(t,{\bf x},{\bf p})$ that coincides with $f$ in the initial state~\cite{Heiselberg:1998es,Borghini:2010hy,Romatschke:2018wgi}: 
\begin{equation}
\label{f_f.s.}
f_{\rm f.s.}(t,{\bf x},{\bf p}) = f^{(0)\!}({\bf x}-{\bf v}t,{\bf p}),
\end{equation}
where ${\bf v}\equiv {\bf p}/|{\bf p}|$ while $f^{(0)}({\bf x},{\bf p})$ denotes the initial distribution (at $t=0$), to which we come back hereafter.

In the second line of Eq.~\eqref{d_tv_n}, $\partial_t N(t)$ is of order ${\cal O}(\sigma)$ (or higher). 
In absence of initial anisotropic flow in the system, $v_n(t)$ is also of order ${\cal O}(\sigma)$, so that the whole term is at least quadratic in $\sigma$: accordingly, we shall neglect it hereafter. 
Note however that this term contributes at linear order in $\sigma$, and thus may not be dropped, if there is some anisotropic flow in the initial state. 

All in all, we replace the evolution equation~\eqref{d_tv_n} with
\begin{equation}
\partial_t v_n(t) = \frac{1}{N}\!\int\!\frac{{\cal C}[f_{\rm f.s.}(t,{\bf x},{\bf p})]}{E_{\bf p}}
  \cos(n \varphi_{\bf p})\,{\rm d}^2{\bf x}\,{\rm d}^2{\bf p},
\label{d_tv_n_O(sigma)}
\end{equation}
valid to linear order in $\sigma$, irrespective of the choice of collision term --- as long as the latter is ${\cal O}(\sigma)$. 
Inserting the loss term~\eqref{loss-term} as collision kernel and integrating over time yields 
\begin{align}
v_n(t) &= -\frac{\sigma}{2N}\!\int_0^t\!\int\!
  f_{\rm f.s.}(t',{\bf x},{\bf p})_{}f_{\rm f.s.}(t',{\bf x},{\bf p}_1) \cos(n \varphi_{\bf p}) \cr
  &\qquad\qquad\quad\times [1-\cos(\varphi_{\bf p}-\varphi_1)] 
  \,{\rm d}^2{\bf x}\,{\rm d}^2{\bf p}\,{\rm d}^2{\bf p}_1\,{\rm d}t' \cr 
 &\quad+ {\cal O}(\sigma^2) \cr
 &\equiv \int_0^t\!\int_0^\infty\!D_n(t',r)\,r\,{\rm d}r\,{\rm d}t' + 
 {\cal O}(\sigma^2),
 \label{v_n(t)_O(sigma)}
\end{align}
where the last line defines the angle-averaged local production rate of $v_n$~\cite{Kurkela:2021ctp}, which we shall discuss in Sect.~\ref{ss:Dn}.
Note that in these expressions we explicitly assumed $v_n(t=0)=0$ in the initial state. 

In our analytical approach the flow coefficients~\eqref{v_n(t)_O(sigma)} depend directly on the initial phase space distribution $f^{(0)}$ via Eq.~\eqref{f_f.s.}.
Let us now discuss our choice for the latter, both for the analytical calculations and the numerical simulations.
First, we assume that the initial phase space distribution factorizes into the product of the particle number density, which determines the geometry, and a position-independent momentum distribution:
\begin{equation}
\label{f^(0)=F.G}
f^{(0)}({\bf x},{\bf p}) = F({\bf x})_{}G({\bf p}),
\end{equation}
where we assume that $G$ is normalized to unity when integrating over the whole two-dimensional momentum space.
This factorization assumption makes our analytical calculations tractable, and enables us to derive analytical formulas for the flow coefficients for the geometrical profile~\eqref{eq:distribution_function}. 
As we shall discuss again in the following, the assumption is however not innocuous, especially for the odd flow harmonics. 
We take $G$ to be isotropic in momentum space, to ensure the absence of initial anisotropic flow. 
Departure from this assumption can be accounted for rather easily, by introducing a Fourier expansion of $G({\bf p})$~\cite{Borghini:2011qc}, but leads to lengthier expressions for the flow coefficients --- whose evolution at linear order in $\sigma$ is no longer governed by Eq.~\eqref{d_tv_n_O(sigma)} as mentioned above.

In position space, we choose as initial density a distorted Gaussian distribution\footnote{In Appendix~\ref{app:different_dist} we briefly present results using an alternative initial density.}
\begin{equation}
F(r,\theta) = \frac{N{\rm e}^{-r^{2\!}/2R^2}}{2\pi R^2}
  \Bigg[1-\sum_{j=2}^6\tilde{\varepsilon}_j {\rm e}^{-r^{2\!}/2R^2} \bigg(\frac{r}{R}\bigg)^{\!\!j} \cos(j\theta)\Bigg],
\label{eq:distribution_function}
\end{equation}
with $N$ the number of particles and $R$ the typical system size, in units of which we shall measure lengths or time.
This form or closely related ones was used extensively in recent studies~\cite{Kurkela:2018qeb,Borghini:2018xum,Kersting:2018qvi,Kurkela:2019kip,Kurkela:2020wwb,Kurkela:2021ctp,Ambrus:2021fej}, as it allows one to introduce at will in the initial state different and independent types of ``eccentricities''~\cite{Alver:2010gr,Teaney:2010vd,Gardim:2011xv}
\begin{align}
\label{eccentricity}
\varepsilon_n {\rm e}^{{\rm i}n\Phi_n}\equiv 
-\frac{\langle r^n {\rm e}^{{\rm i}n\theta}\rangle}{\langle r^n\rangle},
\end{align}
where the angular brackets stand for an average over the transverse plane with some weight, which in the present paper will be the particle-number density. 
Equation~\eqref{eq:distribution_function} yields at once $\Phi_n=0$ --- which we may assume without loss of generality since we shall always consider only a single non-zero $\varepsilon_n$ at a time --- and
\begin{equation}
\varepsilon_n = \frac{(n-1)!}{2^{\frac{2+n}{2}} \Gamma(\frac{n}{2})}_{} \tilde{\varepsilon}_n,
\end{equation}
that is for the first harmonics $\varepsilon_2 = \tilde{\varepsilon}_2/4$, $\varepsilon_3 = \tilde{\varepsilon}_3/\sqrt{2\pi}$, $\varepsilon_4 = 3_{}\tilde{\varepsilon}_4/4$, and so on.
Note that the parameters $\tilde{\varepsilon}_n$ should not be too large, to ensure that the phase space distribution remains non-negative: typically, in case only a single eccentricity is considered, $\tilde{\varepsilon}_n$ should be such that $\varepsilon_n$ remains smaller than $\varepsilon_{n,\text{max}}\simeq 0.35$. 
In our calculations, both analytical and numerical, we choose $\tilde{\varepsilon}_n$ such that $\varepsilon_n = 0.15$ or smaller.

\subsection{Numerical simulations}
\label{ss:numerics}

For our simulations with elastic binary rescatterings, we use the same implementation of the two-dimensional covariant transport algorithm of Ref.~\cite{Gombeaud:2007ub} as in Ref.~\cite{Roch:2020zdl}, to which we refer for further details. 
Here we just recall that the $N$ massless particles are modeled as $N_{\rm p}$ Lorentz-contracted hard spheres --- or rather hard disks, since they are two-dimensional ---  with radius $(N/N_{\rm p})\sigma/2$, where $\sigma$ is the total cross section of the ``physical'' particles. 
Collisions between test particles are determined by a geometric criterion and the scattering angle is deterministic. 
$N_{\rm p}$ and $\sigma$ are always chosen such that the system remains dilute enough, i.e.\ the mean inter-particle distance is at least one order of magnitude smaller than the mean free path $\ell_{\rm mfp}$. 

For the simulations with the $2\to 0$ collision kernel we use the same transport algorithm as in the $2\to2$ case with small modifications.
We introduce labels ``active'' and ``inactive'' for each test particle, such that a collision can only take place between two ``active'' particles, after which they become ``inactive'' and are no longer propagated for the remainder of the simulation.
Eventually, observables like the anisotropic flow coefficients are determined with the ``active'' particles only.

An important difference between this $2\to 0$ model and the analytical approach is that the phase-space distribution in the simulations is affected by rescatterings, i.e.\ the transport simulations include all orders in the cross section.
Thus, we may depart from the few-collision regime in the simulations and investigate what happens when most of the particles disappear due to rescatterings.

Eventually, we also consider a third variant, which we shall refer to as ``single hit'' model, in which particles scatter with the $2\to 2$ kernel, but may undergo at most one collision. 
That is, after their first rescattering --- and the corresponding change in the momenta of the two participants ---, particles become ``transparent'' and stream freely through the system. 
The difference with the $2\to 0$ scenario is that all particles are now taken into account when computing anisotropic flow, irrespective of whether they have undergone zero or one collision. 

In Ref.~\cite{He:2015hfa} the authors used a similar approach with $2\to 2$ and $2\to 0$ collisions. 
The difference to our $2\to 0$ model is that in their study, particles that underwent a collision are still ``active'', but after each collision their momentum azimuths are randomized.
Thus, these particles do indirectly contribute to the generation of anisotropic flow in the azimuth-randomized version of AMPT~\cite{He:2015hfa}.

A crucial ingredient for the comparison with our analytical calculations is the preparation of the initial state of the numerical simulations. 
The test particle positions are sampled from the distribution function~\eqref{eq:distribution_function}, while for their momenta we use a Boltzmann distribution with a position-independent temperature --- in contrast to Ref.~\cite{Borghini:2022qha}.
Since the simulations are performed with a finite test particle number $N_{\rm p}$ ranging between $2\times 10^5$ and $2\times 10^6$, neither perfect isotropy in momentum space nor uniformity of the momentum distribution across the whole geometry can be achieved. 
To improve the situation, for each initial geometry we perform $N_{\rm iter.}$ iterations in which the particles keep the same position but with a different realization of the momentum distribution. 
The results we present are averaged over these iterations, which is expected to diminish fluctuations by a factor $\sqrt{N_{\rm iter.}}$. 
Since the simulation time grows with $N_{\rm p}^{3/2}$, performing multiple iterations with less test particles is computationally less costly than performing a single simulation with $N_{\rm iter.}N_{\rm p}$ particles.\footnote{In our simulations, $N_{\rm iter.}N_{\rm p}$ is always larger than $10^9$.}

\begin{figure}[t]
	\includegraphics*[width=\linewidth]{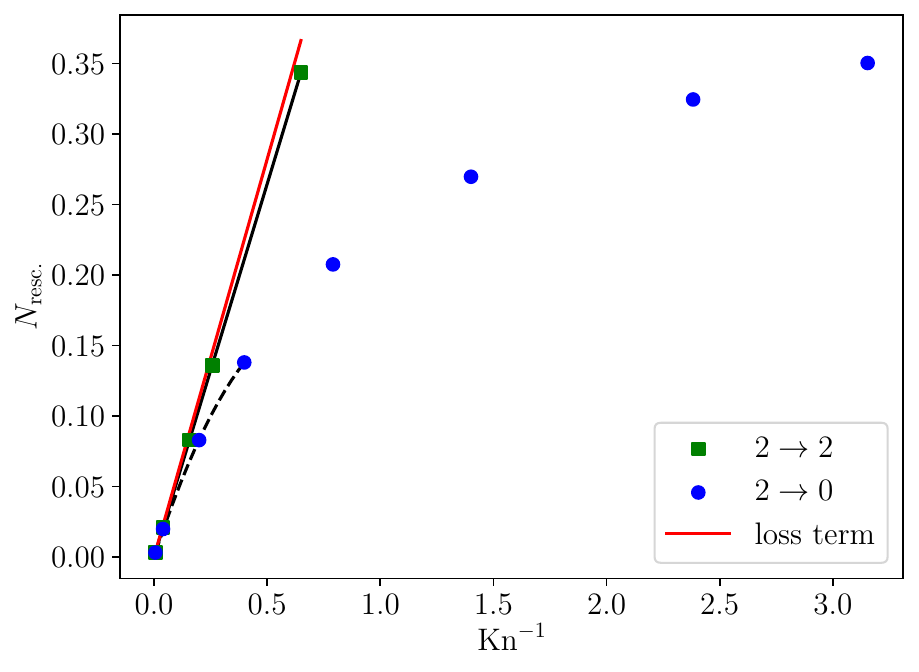}\vspace{-3mm}
	\caption{\label{fig:Nresc_vs_Kn}Mean number of rescatterings per particle over the system evolution as a function of the inverse Knudsen number estimated in the initial state, Eq.~\eqref{Kn_def}, for the $2\to 2$ (green squares, fit with $\Nresc\approx 0.529\,{\rm Kn}^{-1}$) and $2\to 0$ (blue circles, fitted with a quadratic ansatz: dashed line) scenarios. 
		The red line $\Nresc = {\rm Kn}^{-1}/\sqrt{\pi}$ is the prediction of the analytical approach.}
\end{figure}
Starting from Eq.~\eqref{eq:distribution_function}, the average particle-number density per unit surface is $N/4\pi R^2$. 
Using the latter to define a mean free path $\ell_{\rm mfp}$, we quantify the rarity or abundance of rescatterings by the Knudsen number
\begin{equation}
\label{Kn_def}
{\rm Kn}\equiv \frac{\ell_{\rm mfp}}{R} = \frac{4\pi R}{N\sigma},
\end{equation}
with the help of which we shall express the equations resulting from the analytical calculations. 
In contrast, the results of numerical simulations will be presented not at fixed Kn, but rather at fixed mean number of rescatterings per particle $\Nresc$ over the whole evolution --- in practice, until $t/R = 30$. 
We shall mostly present results for $\Nresc\approx 0.02$, well in the few-rescatterings regime, and 0.14 --- for which the approximation becomes less justified ---, as well as $\Nresc\approx 0.35$ in Appendix~\ref{app:Nresc=0.35}.

In the $2\to 2$ scenario, $\Nresc$ nicely scales with ${\rm Kn}^{-1}$, see Fig.~\ref{fig:Nresc_vs_Kn}. 
Note that there are slightly less (about 8\%) rescatterings in our simulations than what would be expected analytically. 
This is due to the finite time step of the transport code, and to the fact that a given particle is allowed to scatter only once per time step, so that we miss collisions,\footnote{We checked that one can capture more collisions by decreasing the time step, which obviously means an increase of computing time.} mostly in the densest regions of the system. 
That is, the effective ${\rm Kn}^{-1}$ in the simulations is actually smaller than that computed from the input parameters, which is a first motivation for presenting numerical results in terms of $\Nresc$ instead.
A second reason for using the mean number of rescatterings per particle is that it turns out that it is the correct scaling variable for comparing systems in the $2\to 0$ scenario, as will be discussed hereafter in Sect.~\ref{ss:v2} and \ref{ss:v3}.

\begin{figure}[t]
\includegraphics[width=\linewidth]{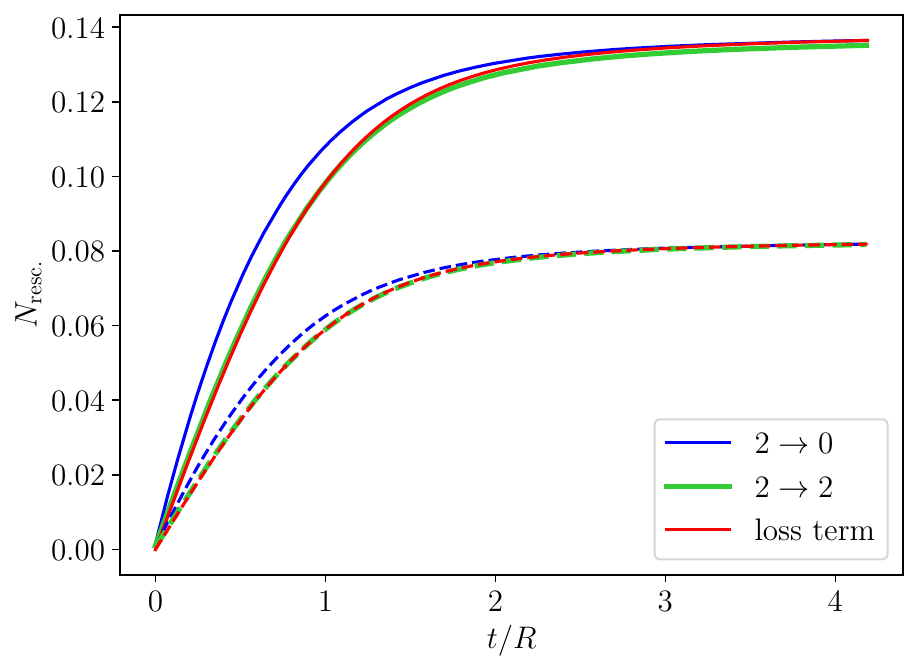}\vspace{-3mm}
\caption{\label{fig:Nresc_vs_t}Time dependence of the cumulative number of rescatterings per particle for systems with in total $\Nresc\approx 0.14$ (full) or 0.08 (dashed) at large times, for simulations with the $2\to 2$ (green) and $2\to 0$ (blue) models, and using Eq.~\eqref{Nresc_vs_t} (red).}
\end{figure}
Nevertheless, it is clear that a given $\Nresc$ requires a larger ${\rm Kn}^{-1}$, i.e.\ cross section, in the $2\to 0$ and single-hit models than in the $2\to 2$ simulations, since particles can never scatter twice in those scenarios.
This in turn means that the collisions tend to occur earlier in the $2\to 0$ and single-hit simulations than in the $2\to 2$ model, as is illustrated in Fig.~\ref{fig:Nresc_vs_t} for azimuthally symmetric systems with in total $\Nresc\approx 0.08$ (dashed) or 0.14 (full lines) collisions per particle.
Accordingly, the geometry of the system at the time of the rescatterings varies across the setups. 
For instance, since the initial asymmetries in the geometry relax as the system expands, one may expect that at the time when anisotropic flow develops --- say roughly for $t/R\leq 2$ --- the system is somewhat more isotropic in the $2\to 2$ simulations than in the other ones, which impacts the anisotropic flow coefficients. 
Anticipating on our findings, this effect does not seem to play a major role. 

In Fig.~\ref{fig:Nresc_vs_t} we also show the time dependence of the number of rescatterings within the analytical approach of Sect.~\ref{ss:analytical}, i.e.\ using the free-streaming phase-space distribution all along the evolution. 
For the initial distribution~\eqref{eq:distribution_function} with vanishing eccentricities one finds
\begin{equation}
\label{Nresc_vs_t}
N_{\rm resc}(t) = \frac{{\rm Kn}^{-1}}{2}\frac{t}{R}_{}
  {\rm e}^{-t^{2\!}/2R^2} \bigg[I_0\bigg(\frac{t^2}{2R^2}\bigg) + I_1\bigg(\frac{t^2}{2R^2}\bigg)\bigg]
\end{equation}
with $I_0$ and $I_1$ modified Bessel functions of the first kind. 
Choosing the value of ${\rm Kn}^{-1}$ such that it yields the same final $\Nresc$ as in the numerical simulations, we see that this formula gives an extremely good approximation to the results in the $2\to 2$ model.

\begin{figure*}[tb!]
\includegraphics[width=0.495\textwidth]{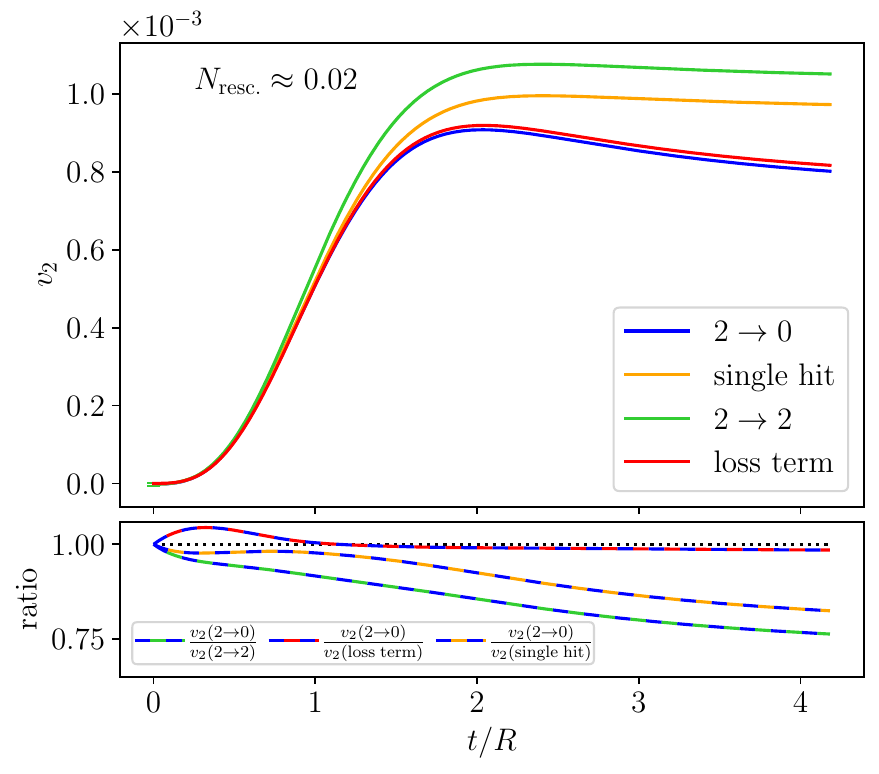}
\includegraphics[width=0.495\textwidth]{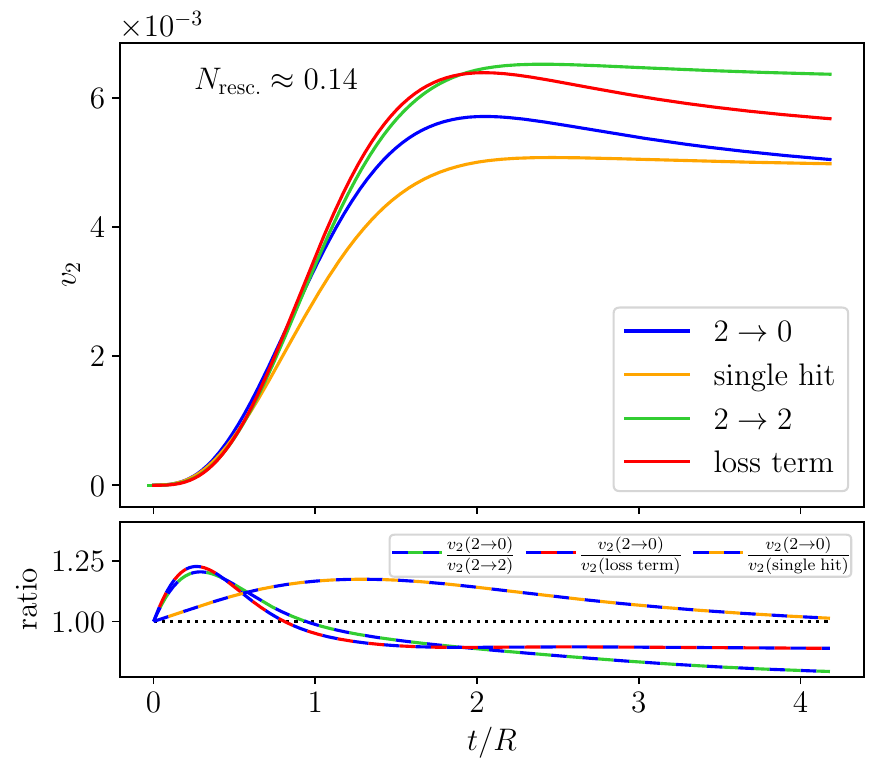}\vspace{-3mm}
\caption{Top: Time dependence of elliptic flow $v_2$ in systems with initially $\varepsilon_2=0.15$ and on average $\Nresc\approx 0.02$ (left) or $0.14$ (right) rescatterings per particle.
The green curves are for systems with elastic binary scatterings, the blue lines for the $2\to 0$ scenario, the orange curves for the single-hit model, and the red ones show the analytical result~\eqref{eq:v2(t)}.
The bottom panels show ratios of the curves from the upper panels.}
\label{fig:v2}
\end{figure*}

\section{Results}
\label{s:Results}

In this Section we present our results for the flow harmonics $v_2$, $v_3$, $v_4$, and $v_6$ for systems with the initial geometry~\eqref{eq:distribution_function}. 
Calculations with a slightly different initial profile, whose results are in qualitative agreement with the findings of this Section, are given in Appendix \ref{app:different_dist}.

\subsection{Elliptic flow}
\label{ss:v2}

Let us start with elliptic flow $v_2$~\cite{Ollitrault:1992bk}. 
As initial geometry we consider the profile~\eqref{eq:distribution_function} with all $\tilde{\varepsilon}_j = 0$ except for $\tilde{\varepsilon}_2$, chosen such that $\varepsilon_2 = 0.15$ (up to numerical fluctuations in the simulations).
The time dependence of $v_2$ in the transport approach is shown in Fig.~\ref{fig:v2} within the $2\to 2$ (green), $2\to 0$ (blue) and single-hit (orange) models, for $N_{\rm resc}\approx 0.02$ (left panel) and $N_{\rm resc}\approx 0.14$ (right panel).%
\footnote{Results in systems with $\Nresc \approx 0.35$ are shown in Fig.~\ref{fig:v2bis}.}
At $t=0$ we indicate as an error bar the typical value $1/\sqrt{2N_{\rm iter.}N_{\rm p}}$ of $v_2$ induced by numerical fluctuations in the initial state. 
We also show in red the result from the analytical calculation, namely
\begin{align}
v_2(t) = \frac{8}{27}_{}{\rm Kn}^{-1\,}\varepsilon_{2\,} {\rm e}^{-2t^2/3R^2}
&\bigg[\!\bigg(\!\frac{3R}{t}+\frac{2t}{R}\!\bigg)I_1\bigg(\!\frac{2t^2}{3R^2}\!\bigg) \cr 
&\quad-\frac{t}{R}I_0\bigg(\!\frac{2t^2}{3R^2}\!\bigg)\bigg]
\label{eq:v2(t)}
\end{align}
where the value of ${\rm Kn}^{-1}$ is chosen such that it gives the same $\Nresc$ as in the numerical calculations.
Note that Eq.~\eqref{eq:v2(t)} yields $v_2(t)\propto t^3$ at early times $t\ll R$, as pointed out in previous studies~\cite{Gombeaud:2007ub,Alver:2010dn,Borghini:2010hy,Borrell:2021cmh}.

To quantify the deviation between the various approaches, we fitted our results from transport simulations, shifted to $v_2(t=0) = 0$ for a better comparison, with respective Pad\'e approximants
\begin{equation}
v_2(t) \sim \frac{\sum_{k=3}^{5} a_k (t/R)^k}{1+\sum_{k=1}^{5} b_k (t/R)^k}
\label{eq:Pade}
\end{equation}
to wash out the numerical fluctuations, especially at early times.
A drawback from the approximation is that the fits are dominated by the values for $t/R\gtrsim 1$, so that the early time behaviors are not necessarily captured correctly. 
Using these fits, we computed the ratios of the $v_2$ values in the $2\to 0$ scenario either to those of the $2\to 2$ and single-hit models or to the analytical value~\eqref{eq:v2(t)}, and show these ratios in the narrow lower panels in Fig.~\ref{fig:v2}. 

The profiles of $v_2(t)$ are similar in the four approaches, with a slow onset, followed by an almost linear rise, that eventually saturates. 
$v_2$ reaches its maximum value for $t/R \approx 2$, and decreases a little afterwards, barely in the $2\to 2$ and single-hit scenarios.
Remarkably, the overall shape of $v_2(t)$ is the same for the small numbers of rescatterings considered here as in the fluid-dynamical limit, illustrated e.g.\ in Ref.~\cite{Alver:2010dn} (Fig.~3, with a slightly different geometry).

\begin{figure}[t]
	\includegraphics*[width=\linewidth]{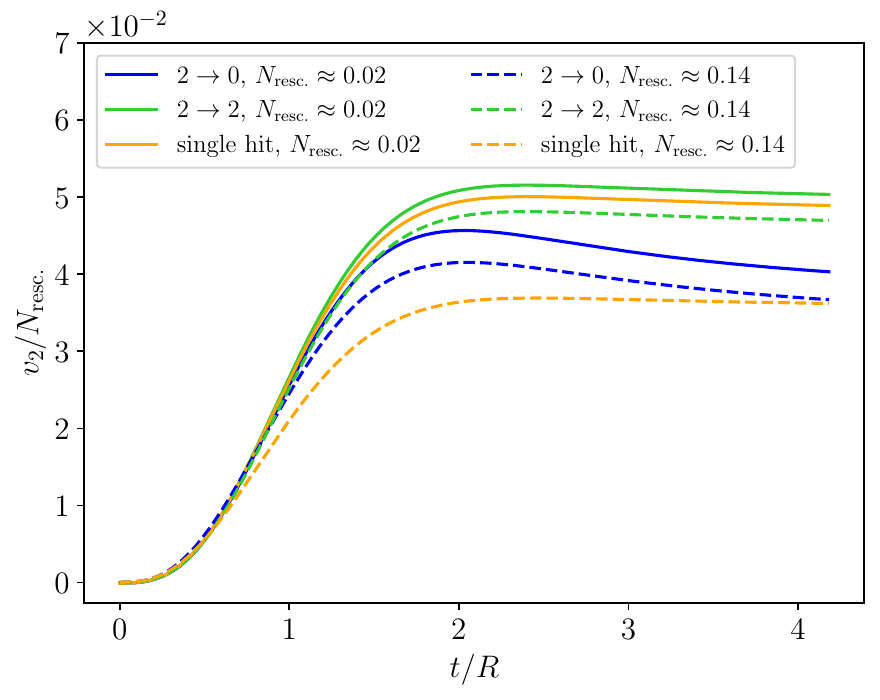}\vspace{-3mm}
	\caption{\label{fig:v2/Nresc}Ratio $v_2(t)/\Nresc$ in systems with initially $\varepsilon_2=0.15$ and on average $\Nresc\approx 0.02$ (full lines) or $0.14$ (dashed) rescatterings per particle for the three scenarios of the transport cascade: 
		$2\to 2$ (green), $2\to 0$ (blue), single hit (orange).}
\end{figure}

\begin{figure*}[t!]
	\includegraphics[width=0.495\linewidth]{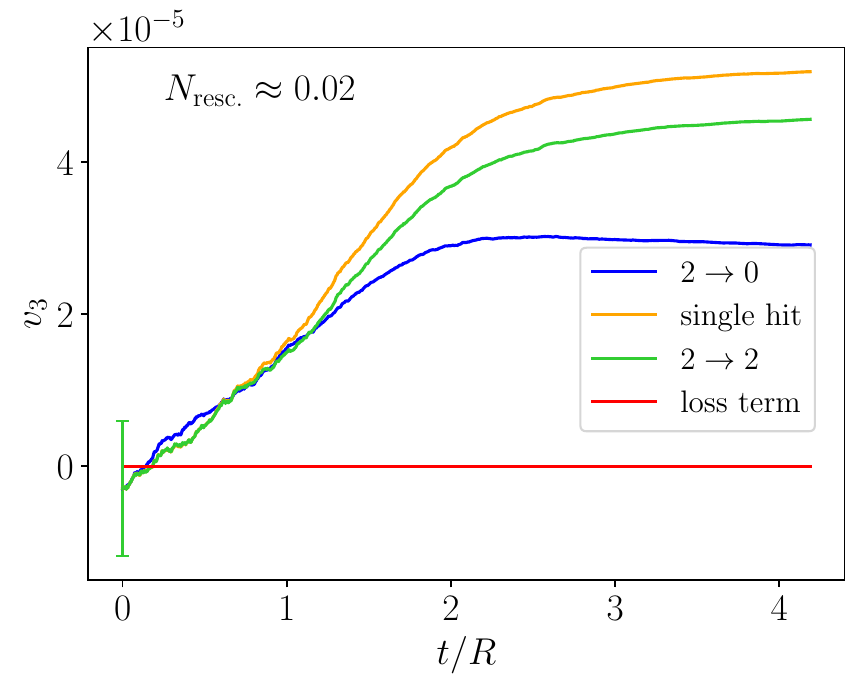}
	\includegraphics[width=0.495\linewidth]{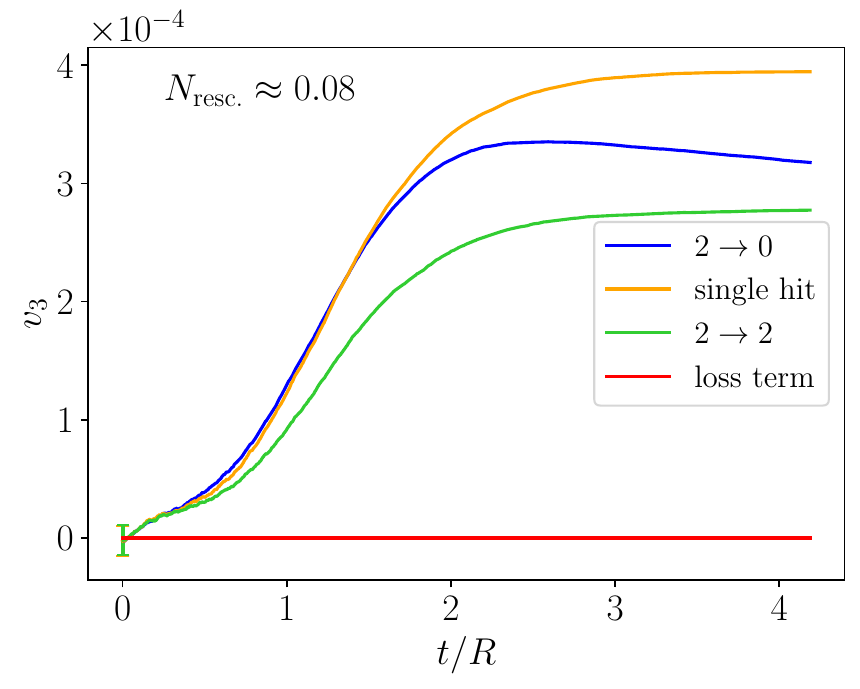}
	\includegraphics[width=0.495\linewidth]{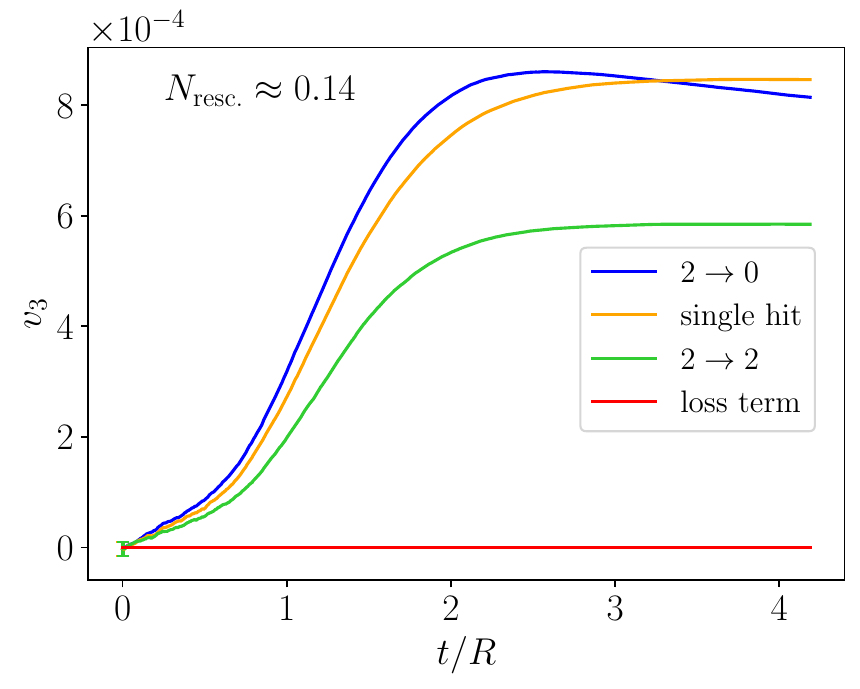}\vspace{-3mm}
	\caption{Time dependence of triangular flow $v_3$ in systems with initially $\varepsilon_3=0.15$ and on average $\Nresc\approx 0.02$ (top left), 0.08 (top right) or $0.14$ (bottom) rescatterings per particle.
		The green curves are for systems with elastic binary scatterings, the blue lines for the $2\to 0$ collision kernel, and the orange curves for the single-hit model. 
		The constant red line $v_3=0$ is the output of the analytical approach.}
	\label{fig:v3}
\end{figure*}

More importantly for the purpose of this paper, the elliptic flow built up in the $2\to 0$ model differs at most by 20\% from that in the ``full'' $2\to 2$ case. 
In addition, the results of the $2\to 0$ scenario are extremely well reproduced by the analytical calculations accounting only for the loss term for $\Nresc\approx 0.02$.  
The agreement is less impressive but still very good at the larger $\Nresc$, which is easily understood: 
The analytical results are derived at linear order in the cross section, or equivalently $\mathrm{Kn}^{-1}$. 
As the latter increases, higher order contributions to $v_2$, which are always present in the $2\to 0$ simulations, become more sizable, and lead to the departure between the analytical results and the $2\to 0$ values.
Indeed, we have shown in Ref.~\cite{Borghini:2022qha} --- yet only for early times --- that pushing the analytical calculation to higher order in $\sigma$ improves the agreement with the $2\to 0$ results.
In contrast, for $t/R\geq 2$, when fewer collisions take place, the results of both approaches are again very parallel. 

Eventually, the results from the single-hit scenario show a non-systematic trend with varying $\Nresc$. 
When the number of rescatterings is very small ($\Nresc\approx 0.02$), the single-hit $v_2(t)$ is intermediate between the $2\to 0$ and $2\to 2$ results. 
This seems consistent with the intuition that the single-hit model captures part of the gain term of the Boltzmann equation --- since colliding particles are redistributed in momentum space ---, but not the whole of it, as particles can scatter at most once. 
However, when $\Nresc$ increases, the single-hit results for $v_2(t)$ depart more strongly from those of the $2\to 2$ cascade, and they are now further away from them as those from the $2\to 0$ scenario, see right panel of Fig.~\ref{fig:v2}. 

The somewhat different behavior of the single-hit model, for which we could not find an easy explanation, is also illustrated in Fig.~\ref{fig:v2/Nresc}, which displays $v_2(t)$ scaled by the total number of rescatterings for the three scenarios of our transport code and for the two values $\Nresc\approx 0.02$ and $0.14$. 
This figure shows that to a very good approximation $v_2(t) \propto \Nresc$ holds in the full $2\to 2$ simulations ---- it is then equivalent to $v_2\propto{\rm Kn}^{-1}$, see Fig.~\ref{fig:Nresc_vs_Kn}, i.e.\ $v_2\propto\sigma$ --- and the $2\to 0$ model, but the scaling is less good, although still satisfactory, for the single-hit case. 
In Fig.~\ref{fig:v2/sigma} in Appendix~\ref{app:v2_vs_sigma} we show for the sake of completeness the ratio of $v_2(t)/{\rm Kn}^{-1}$, i.e.\ essentially of elliptic flow over the cross section, for the same simulations as in Fig.~\ref{fig:v2/Nresc}. 
For the $2\to 0$ and single-hit scenarios, the curves corresponding to systems with $\Nresc\approx 0.02$ and $0.14$ are far apart from each other, which shows that $\Nresc$ is indeed a better scaling variable than the inverse Knudsen number ${\rm Kn}^{-1}$ for those simulations (at least as far as anisotropic flow is concerned).

All in all, we find that in the few-rescatterings regime 
most of the $v_2$ signal may be ascribed to the processes modeled by the loss term of the Boltzmann equation.
That is, the elliptic flow in the final state seems to arise to a large extent from the anisotropic survival probability of the particles as they propagate through the system~\cite{Heiselberg:1998es,Borghini:2010hy}, as advocated in the ``escape mechanism'' picture~\cite{He:2015hfa}.

\subsection{Triangular flow}
\label{ss:v3}

\begin{figure*}[t]
	\includegraphics[width=0.495\linewidth]{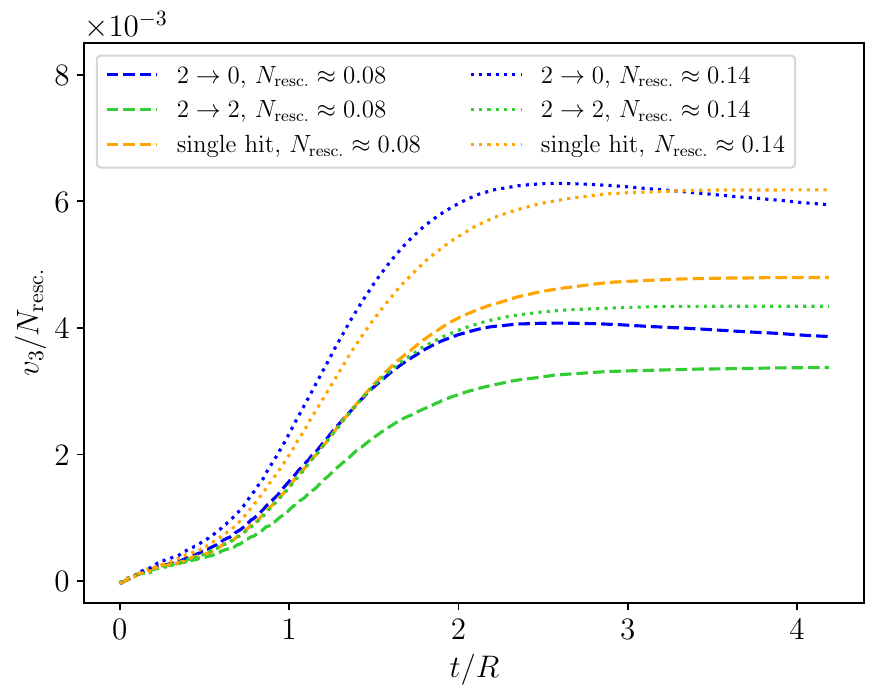}
	\includegraphics[width=0.495\linewidth]{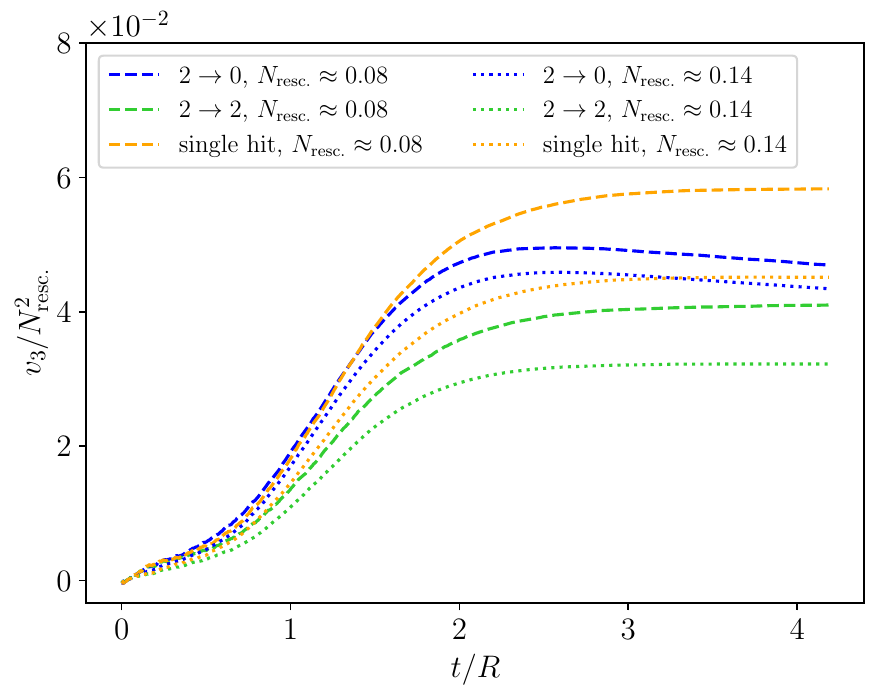}\vspace{-3mm}
	\caption{\label{fig:v3/Nresc}Ratios $v_3(t)/\Nresc$ (left) and $v_3(t)/\Nresc^2$ (right) in systems with initially $\varepsilon_3=0.15$ and on average $\Nresc\approx 0.08$ (dashed lines) or $0.14$ (dotted lines) rescatterings per particle for the three scenarios of the transport cascade: 
		$2\to 2$ (green), $2\to 0$ (blue), single hit (orange).}
\end{figure*}

We turn next to triangular flow $v_3$~\cite{Alver:2010gr}, using now an initial geometrical profile~\eqref{eq:distribution_function} with only a non-zero $\tilde{\varepsilon}_3$, such that $\varepsilon_3 = 0.15$.
The results of our various calculations for the time dependence of $v_3$ are displayed in Fig.~\ref{fig:v3}, for systems with $\Nresc\approx 0.02$ (top left), 0.08 (top right) or $0.14$ (bottom).

A first striking feature is that $v_3$ identically vanishes in the analytical approach if it is zero initially.
As we show in Appendix~\ref{app:v3,v5} and discuss again in Sect.~\ref{ss:Dn}, this is due to a cancellation between different regions in the special case --- which we consider throughout the paper --- where the local momentum distribution is the same at every point of the transverse plane in the initial condition. 
To be more precise, one finds that $v_3$, and more generally every odd flow harmonic, is zero at first order in $\sigma$, but at higher orders it can be non-zero~\cite{Borghini:2022qha}. 

As to the results of transport simulations, we see a number of differences with those for elliptic flow. 
First, the $v_3$ signal is an order of magnitude smaller than $v_2$, so that the curves are more affected by the numerical fluctuations, in particular in the initial state.\footnote{The analytical calculation with an initial momentum anisotropy leads to a non-vanishing and slightly evolving $v_3$. The latter is however negligible compared to the values of the numerical simulations and therefore not shown in Fig.~\ref{fig:v3}.}
Secondly, the simulations within the $2\to 0$ model give a clear non-zero signal, in contrast to the analytical result. 
This hints that in the $2\to 0$ simulations, which include all orders in the cross section, $v_3$ arises at a higher order in $\sigma$. 

Thirdly, the results of the $2\to 0$ scenario clearly do not resemble those of the $2\to 2$ model. 
For $\Nresc\approx 0.02$, the $2\to 0$ results lie about a factor 1.5 below, while they are larger for $\Nresc\approx 0.08$ and 0.14.

Eventually, the results of the single-hit model for $v_3(t)$ again show no clear trend in comparison to the other two numerical models. 
At $\Nresc\approx 0.02$ they closely resemble the results of the $2\to 2$ computations --- the overshooting is probably due to the initial noise. 
But at larger number of rescatterings they are closer to the outcome of the $2\to 0$ simulations, which makes it difficult to draw any conclusion. 

In Fig.~\ref{fig:v3/Nresc} we compare systems with different number of rescatterings by scaling $v_3$ by $\Nresc$ (left) or $\Nresc^2$ (right). 
Since the simulations with $\Nresc\approx 0.02$ are largely plagued by noise in the ``early stage'' $t\lesssim R$, we discard them from the comparison and only look at $\Nresc\approx 0.08$ and 0.14.
For the $2\to 0$ scenario, the plots hint at a scaling behavior $v_3\propto\Nresc^2$, different from that found for elliptic flow.  
Regarding the $2\to 2$ (and even more the single-hit models), the plots are rather inconclusive, and both scalings with $\Nresc$ and $\Nresc^2$ seem almost acceptable. 
Let us note that studies focusing on the final value of $v_3$, at the end of the evolution, have found $v_3\propto\Nresc$ (or equivalently $v_3\propto{\rm Kn}^{-1}$) at small $\Nresc$ in systems with elastic binary scalings~\cite{Alver:2010dn,Borghini:2022qha}.

\begin{figure*}[t!]
	\includegraphics[width=0.503\linewidth]{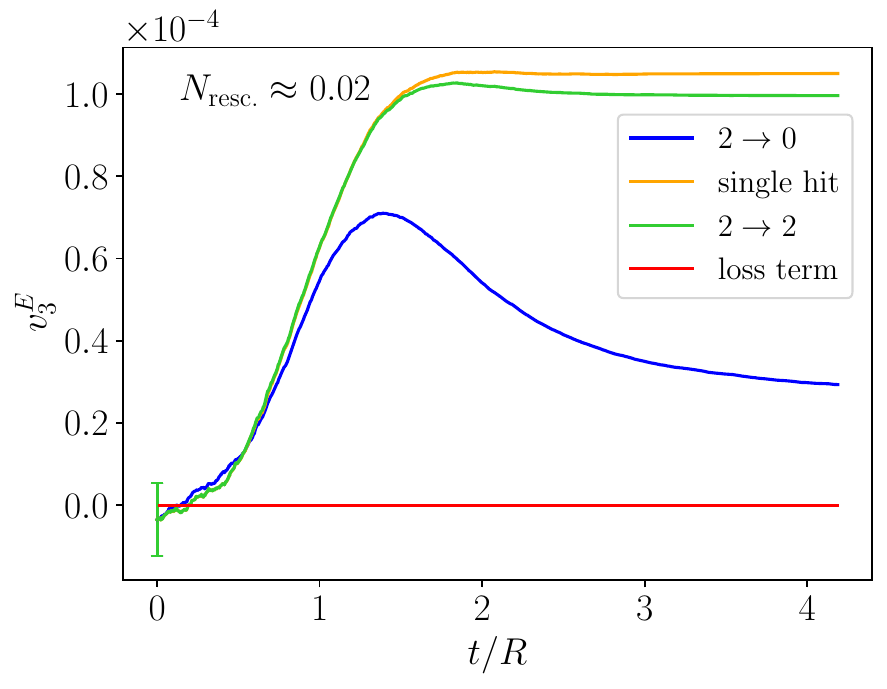}
	\includegraphics[width=0.487\linewidth]{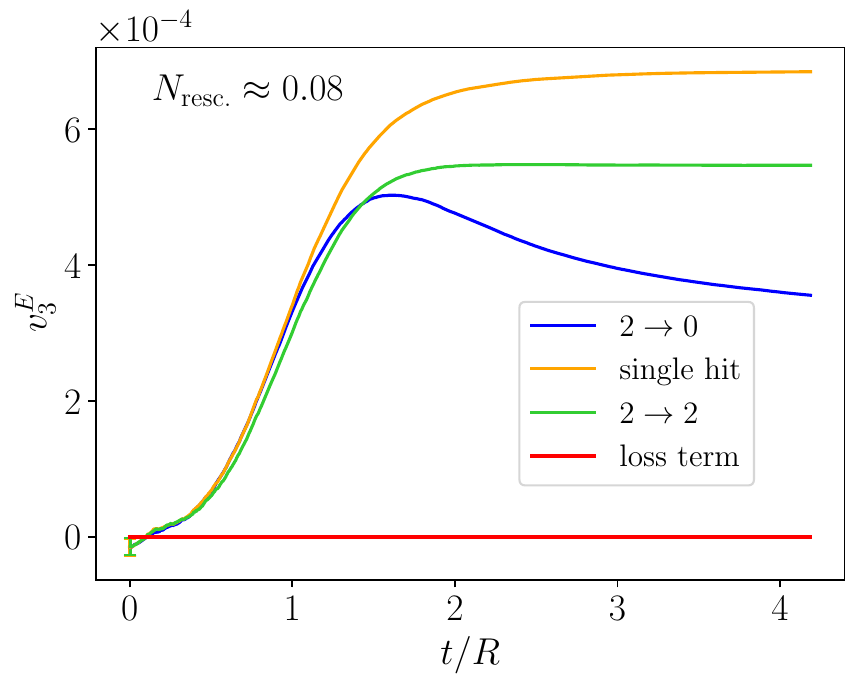}
	\includegraphics[width=0.495\linewidth]{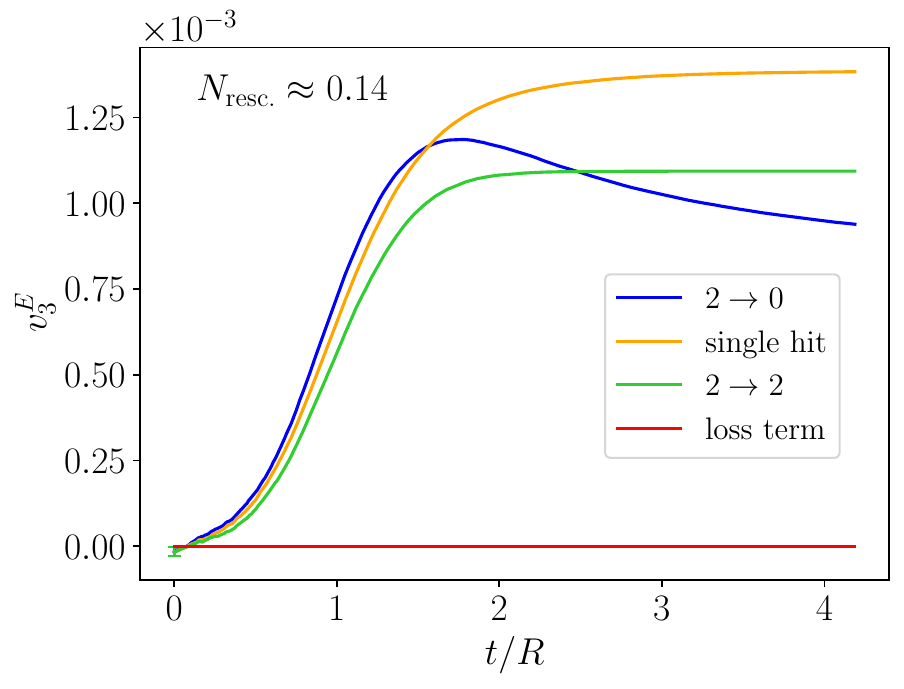}\vspace{-3mm}
	\caption{Time dependence of energy-weighted triangular flow $v_3^E$ in systems with initially $\varepsilon_3=0.15$ and on average $\Nresc\approx 0.02$ (top left), 0.08 (top right) or $0.14$ (bottom) rescatterings per particle.
		The green curves are for systems with elastic binary scatterings, the blue lines for the $2\to 0$ collision kernel, and the orange curves for the single-hit model. 
		The constant red line $v_3^E=0$ is the output of the analytical approach.}
	\label{fig:v3E}
\end{figure*}

Several recent studies investigated the ``energy weighted triangular flow'' $v_3^E$, i.e.\ the third Fourier coefficient of the transverse energy distribution, instead of the particle-number weighted coefficients~\cite{Kurkela:2020wwb,Kurkela:2021ctp,Ambrus:2021fej}. 
As shown in Fig.~\ref{fig:v3E}, $v_3^E$ --- computed in the same systems as used for Fig.~\ref{fig:v3} --- again differs a lot in the $2\to 2$ and $2\to 0$ scenarios. This is especially true at times $t\gtrsim R$. 
In turn, the single-hit results are quite close to the $2\to 2$ values at $\Nresc\approx 0.02$, but at higher $\Nresc$ they tend to be systematically larger.
At earlier times $t\lesssim R$, the results with the three scenarios are more similar, but this is possibly a coincidence, as part of that early behavior is driven by numerical fluctuations: 
due to the finite number of particles, it is impossible to impose that the momentum distribution be exactly isotropic and identical everywhere in the transverse plane, so that the numerical realizations differ from the idealized setup.

In summary, and in strong contrast to the findings of Sect.~\ref{ss:v2}, we find that for $v_3$ the $2\to 0$ scenario differs significantly from the $2\to 2$ model. 
In parallel, the triangular flow from the analytical approach considering only the loss term at first order in $\sigma$ is also at variance with the results of numerical simulations.%
\footnote{In Ref.~\cite{Borghini:2022qha} --- in which a slightly different setup is used, namely with initially a thermal momentum distribution with a position-dependent temperature ---, the results of analytical calculations for $v_3(t)$ at order $\sigma^2$ but restricted to early times $t\leq R$ are found to be of the same magnitude as those of numerical computations, but the shape (which is affected by numerical noise) is not reproduced.}
This is a strong hint that the final state of the individual rescatterings, modeled by the gain term of the Boltzmann equation, plays a crucial role: 
That is, the observed $v_3$ is not carried predominantly by particles that underwent no rescattering and escaped anisotropically from the medium.

\subsection{Quadrangular flow}
\label{ss:v4}

With quadrangular flow $v_4$, the situation is again simpler than for $v_3$. 
Anticipating on what we shall now present, the overall trend is the same as for elliptic flow $v_2$: 
the results of the numerical $2\to 2$, $2\to 0$ and single-hit simulations and those of the analytical approach nicely agree when the number of rescatterings is (very) small, hinting at the dominant role of the escape mechanism for $v_4$ in this regime. 
\begin{figure*}[tb!]
\includegraphics[width=0.495\linewidth]{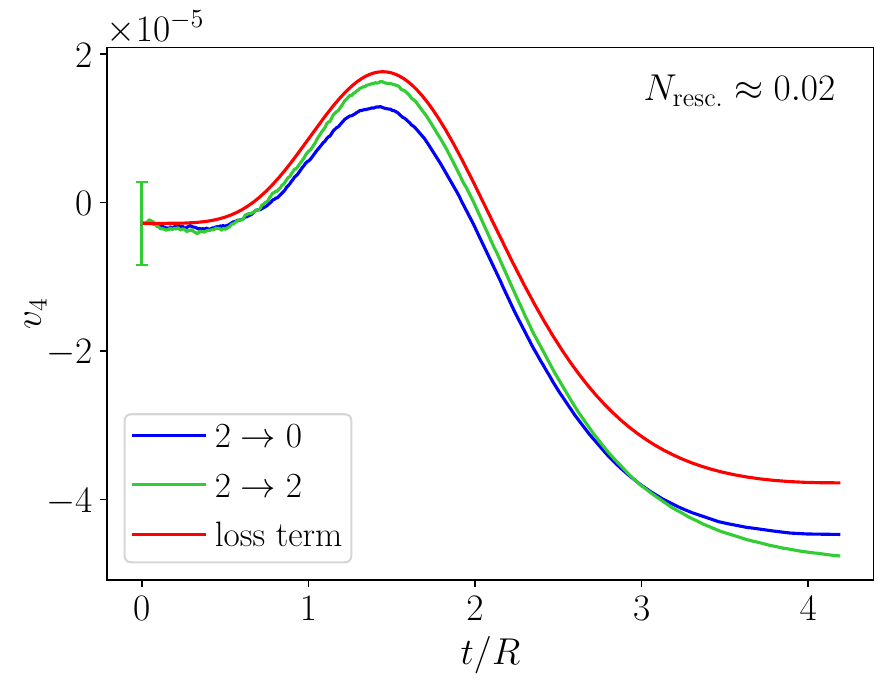}
\includegraphics[width=0.495\linewidth]{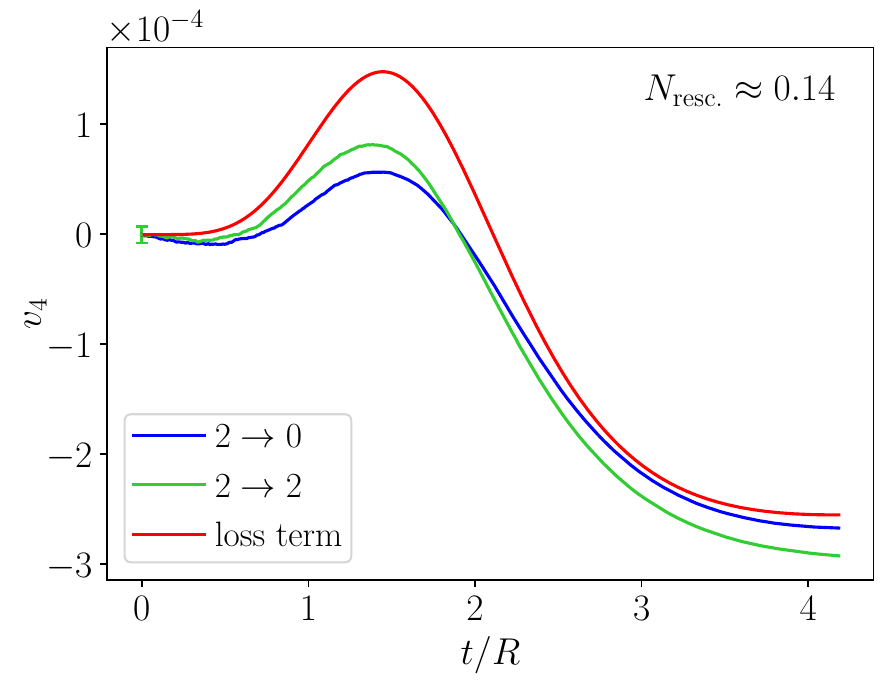}
\includegraphics[width=0.495\linewidth]{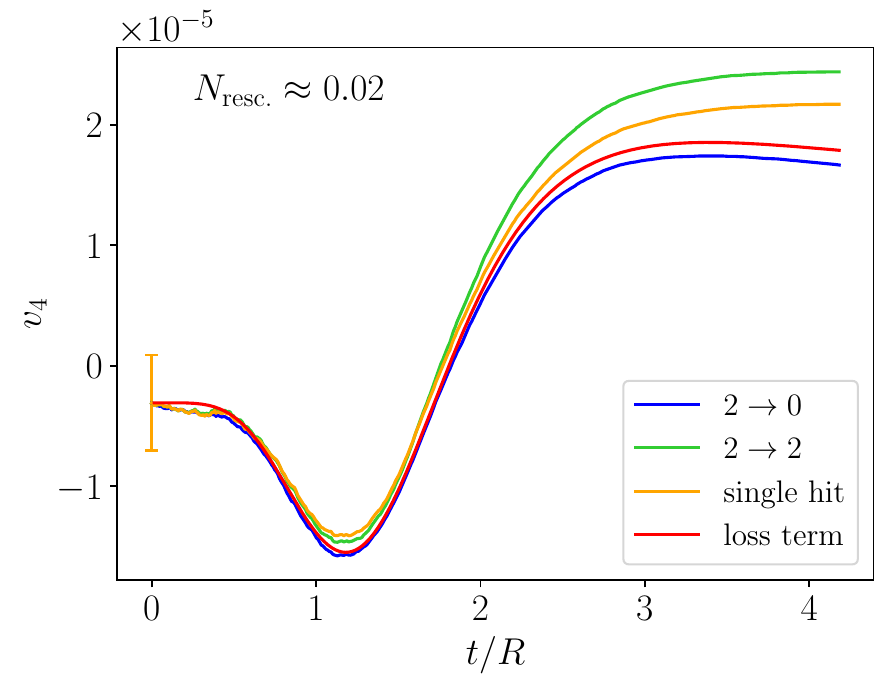}
\includegraphics[width=0.495\linewidth]{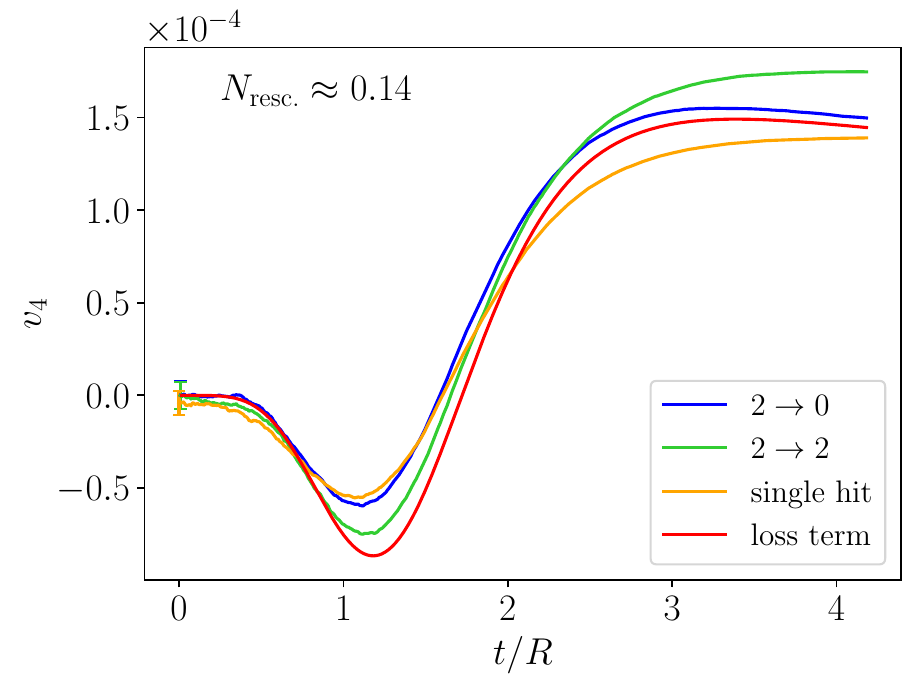}\vspace{-3mm}
\caption{Time dependence of quadrangular flow $v_4$ in systems with on average $\Nresc\approx 0.02$ (left) or $0.14$ (right) rescatterings per particle.
The top panels are for a system with an initial quadrangularity  $\varepsilon_4 = 0.15$ and $\varepsilon_2 = 0$; 
the bottom panels for an initial state with $\varepsilon_2 = 0.15$ and $\varepsilon_4 = 0$.
Green: systems with elastic binary scatterings; blue: $2\to 0$ kernel; orange: single-hit model; red: analytical results~\eqref{eq:v4_e4(t)} (top) or Eq.~\eqref{eq:v4_e2(t)} (bottom).}
\label{fig:v4}
\end{figure*}

Starting with $v_4$, a new possibility appears, namely that the produced anisotropic flow harmonic $v_n$ can arise not only because of the spatial harmonic $\varepsilon_n$, but also due to nonlinear effects mixing other eccentricities.%
\footnote{To be more accurate, according to our present knowledge the lower flow harmonics $v_n$ with $n\leq 3$ are only minimally affected by such nonlinear effects involving eccentricities $\varepsilon_k$ with $k\neq n$.}
Thus, $v_4$ may be caused not only by the ``quadrangularity'' $\varepsilon_4$, but also by the ``ellipticity'' $\varepsilon_2$~\cite{Borghini:2005kd,Gardim:2011xv,Teaney:2012ke,Niemi:2012aj,Borghini:2018xum}.

Indeed, our analytical calculation for $v_4$ assuming only a non-vanishing $\tilde{\varepsilon}_4$ in the initial state yields
\begin{align}
v_4(t) =\ &\frac{16}{1215}_{}{\rm Kn}^{-1\,}\varepsilon_{4\,}{\rm e}^{-2t^2/3R^2} \cr
  &\times \bigg[ \bigg(\frac{162R^3}{t^3}+\frac{63R}{t}+\frac{24t}{R}+\frac{5t^3}{R^3}\bigg)I_1\bigg(\frac{2t^2}{3R^2}\bigg) \cr
  &\qquad - \bigg(\frac{54R}{t}+\frac{21t}{R}+\frac{5t^3}{R^3}\bigg)I_0\bigg(\frac{2t^2}{3R^2}\bigg) \bigg].
\label{eq:v4_e4(t)}
\end{align}
Assuming instead that only a non-vanishing $\tilde{\varepsilon}_2$ is initially present, we obtain
\begin{align}
v_4(t) =\ &\!-\!\frac{1}{10}_{}{\rm Kn}^{-1\,}\varepsilon_{2\,}^2{\rm e}^{-t^2/R^2} \cr
&\times  \bigg[ \bigg(\frac{48R^3}{t^3}+\frac{28R}{t}+\frac{16t}{R}+\frac{5t^3}{R^3}\bigg)I_1\bigg(\frac{t^2}{R^2}\bigg) \cr
&\qquad -   \bigg(\frac{24R}{t}+\frac{14t}{R}+\frac{5t^3}{R^3}\bigg)I_0\bigg(\frac{t^2}{R^2}\bigg) \bigg].
\label{eq:v4_e2(t)}
\end{align}
Obviously, the terms on the right-hand sides of these equations add up if the initial state contains both $\tilde{\varepsilon}_2$ and $\tilde{\varepsilon}_4$.
These analytical results are compared to those of numerical simulations with both $2\to 2$ (green) and $2\to 0$ (blue) collision kernels in Fig.~\ref{fig:v4}: 
the plots in the top panels are with $\tilde{\varepsilon}_4\neq 0$, such that $\varepsilon_4=0.15$, and all other $\tilde{\varepsilon}_k=0$, while the bottom panels --- in which we also show the results from simulations in the single-hit scenario (orange) --- are for a non-zero  $\tilde{\varepsilon}_2$ (with $\varepsilon_2=0.15$) and vanishing other eccentricities.\footnote{These simulations with only an ellipticity $\varepsilon_2=0.15$ are actually the same as used for $v_2$ in Sect.~\ref{ss:v2}.} 
Figure~\ref{fig:v4} displays the time evolution of $v_4$ for systems with $\Nresc\approx 0.02$ (left) or 0.14 (right) rescatterings per particle, while results for $\Nresc\approx 0.35$ are shown in Fig.~\ref{fig:v4bis}.

Overall, the results in Fig.~\ref{fig:v4} show that in the case of quadrangular flow $v_4$, either from $\varepsilon_2$ or from $\varepsilon_4$, the $2\to 0$ model represents a very good approximation of the $2\to 2$ collision kernel for low $\Nresc$.
In turn, the nice agreement with the analytical results reinforces that statement and shows that $v_4$ is proportional to $\sigma$ in that regime. 
Indeed, the less good agreement of the ``loss term'' results with the $2\to 0$ kernel for $\Nresc = 0.14$ can be attributed to the limitation of the analytical calculations to linear order in the cross section. 
At both values of $\Nresc$ and for collisions with an initial ellipticity, the $v_4$ values from the single-hit model also roughly match those of the $2\to 2$ simulations, although less so at the larger $\Nresc$. 

Although the results of Fig.~\ref{fig:v4} suggest that $v_4$ behaves as $v_2$, in that it seems to be mostly driven by the particles that did not collide --- at least in the low $\Nresc$ regime ---, still there are important differences. 
A first one, to which we shall come back in Sect.~\ref{ss:Dn}, is that $v_4$ changes sign over time, while $v_2$ does not. 
A second difference is that while the overall shape of $v_2(t)$ is roughly the same in the few-rescatterings regime and in the fluid-dynamical limit, this does not hold true for $v_4(t)$. 
Indeed, we find that for a larger number of rescatterings ($\Nresc\gtrsim 5$, with the $2\to 2$ collision kernel, since the $2\to 0$ scenario makes no sense in that case) the $v_4$ resulting from an initial $\varepsilon_4>0$ is positive at late times, as found also e.g.\ in Refs.~\cite{Alver:2010dn,Kurkela:2020wwb},\footnote{Strictly speaking, in Ref.~\cite{Alver:2010dn} a different initial profile was used, namely Eq.~\eqref{eq:different_distribution_function}. In turn, the results of Ref.~\cite{Kurkela:2020wwb} are for energy-weighted quadrangular flow $v_4^E$, but we checked that it behaves like $v_4$ in our setup. It seems that our small $\Nresc$ regime is actually beyond the low-opacity region studied in Ref.~\cite{Kurkela:2020wwb}.} but contrary to the behavior of the upper panels of Fig.~\ref{fig:v4}.
This means that the linear scaling with $\Nresc$ of the ``final'' $v_4$ observed in Fig.~\ref{fig:v4} breaks down at larger cross sections. 
Note that a negative $v_4$ --- more accurately, $v_4^E$ --- for a positive $\varepsilon_4$ in the few-rescatterings regime was also found in Ref.~\cite{Ambrus:2021fej}, yet with a different collision kernel based on the relaxation time approximation. 
This difference in the collision kernel may explain why we do not find the same behavior at early times --- namely a negative $v_4$ --- in case the system is initially deformed elliptically ($\varepsilon_2\neq 0$, $\varepsilon_4=0$).

All in all, it seems that in the few-rescatterings regime $v_4$, either resulting ``linearly'' from an initial quadrangularity $\varepsilon_4$ or nonlinearly from an initial ellipticity $\varepsilon_2$, behaves like elliptic flow $v_2$, i.e.\ it largely arises from the anisotropic escape of particles. 
Interestingly, the contributions from $\varepsilon_2$ and $\varepsilon_4$ to $v_4$ are of the same order of magnitude, and in the small $\Nresc$ regime they are of opposite signs.
Accordingly, the two contributions can partly cancel each other and lead to a $v_4$ value at large times that can lie in a wide range of values.
In particular, it is possible to obtain a negative $v_4$ value.

\subsection{Hexagonal flow}
\label{ss:v6}

Going beyond $v_4$, we can guess qualitatively in analogy to our study of $v_3$ what we would find for $v_5$: since it is an odd harmonic, the analytical approach gives zero at linear order in $\sigma$.
In turn, this means that in the $2\to 0$ scenario $v_5$ arises at order $\Nresc^2$, while it is proportional to $\Nresc$ in the $2\to 2$ model, so that we would find discrepancies between the two types of transport simulations. We did not attempt to perform such simulations, which would require new sets of events with the appropriate controlled initial geometry.

\begin{figure}[tb!]
\includegraphics[width=\linewidth]{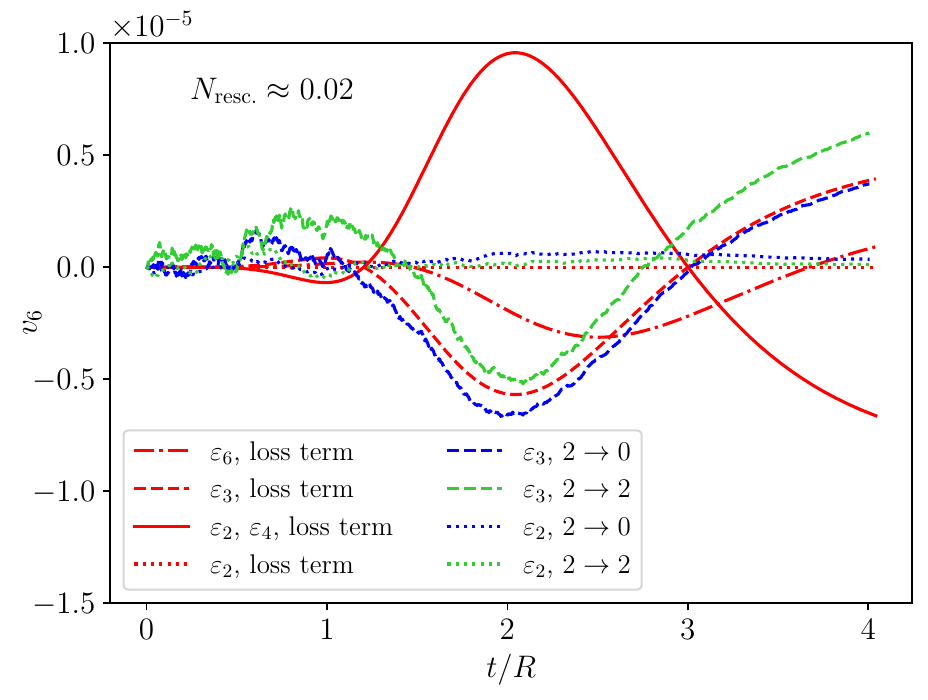}\vspace{-3mm}
\caption{\label{fig:v6}Time dependence of hexagonal flow $v_6$ in systems with on average $\Nresc\approx 0.02$ per particle and with different initial geometrical profiles, as described in the text.}
\end{figure}

Here we present in Fig.~\ref{fig:v6} results for $v_6$, which is at the limit of what we can do numerically with reasonable control on the signal when $\Nresc\approx 0.02$, while exploiting simulations that were already used for $v_2$ or $v_3$.
Indeed, an interesting feature of $v_6$ is that it can result from different initial geometries~\cite{Bravina:2013ora,Qian:2016fpi,Giacalone:2018wpp}, in particular with only a hexagonal deformation (linear response $v_6\propto \varepsilon_6$, dot-dashed line), only an initial triangularity (quadratic response $v_6\propto \varepsilon_3^2$, dashed lines), only an initial $\varepsilon_2$ (cubic response $v_6\propto \varepsilon_2^3$, dotted lines), or with both initial $\varepsilon_2$ and $\varepsilon_4$ (mixed quadratic response $v_6\propto \varepsilon_2\varepsilon_4$, full line).
In every setup the only non-zero $\varepsilon_n$ are set to $0.15$. 
The numerical results with an initial $\varepsilon_2$ resp.\ $\varepsilon_3$ are from the same simulations as in Sect.~\ref{ss:v2} resp.\ \ref{ss:v3}. 
We did not attempt to perform simulations with an initial non-zero $\varepsilon_6$ nor with both $\varepsilon_2$ and $\varepsilon_4$ (and aligned symmetry planes $\Phi_2$ and $\Phi_4$, as assumed for the analytical curve).

Similarly to what we found for $v_2$ and $v_4$, the results for $v_6$ stemming from an initial $\varepsilon_3$ agree rather well across the three scenarios of this paper in the few-rescatterings regime. 
This agreement should be contrasted with Sect.~\ref{ss:v3}, in which the same initial setup yielded very disparate results for $v_3$. 
This reinforces our main conclusion of the paper regarding the different ``origins'' of the even and odd flow harmonics.

As regards the $v_6$ from an initial $\varepsilon_2$, the results from numerical simulations are extremely small but seem to be non-zero and consistent in the $2\to 2$ and $2\to 0$ models. 
In contrast, the analytical results in that case are exactly zero: as was pointed out in Ref.~\cite{Borghini:2018xum}, in a model with only binary collisions and no quantum-statistical effects, a contribution in $\varepsilon_2^3$ to $v_6$ (or to $v_2$) can only arise at order $\sigma^2$, not at linear order in $\sigma$ as considered here. 

Eventually, the analytical results for initial geometries with either $\varepsilon_6=0.15$ or $\varepsilon_2=\varepsilon_4=0.15$ are of the same typical magnitude as those for $\varepsilon_3=0.15$. 
As in the case of $v_4$, the signal changes sign (here twice) over the system evolution.

\subsection{Local production rate of anisotropic flow}
\label{ss:Dn}

\begin{figure*}[tb!]
\includegraphics[width=0.495\textwidth]{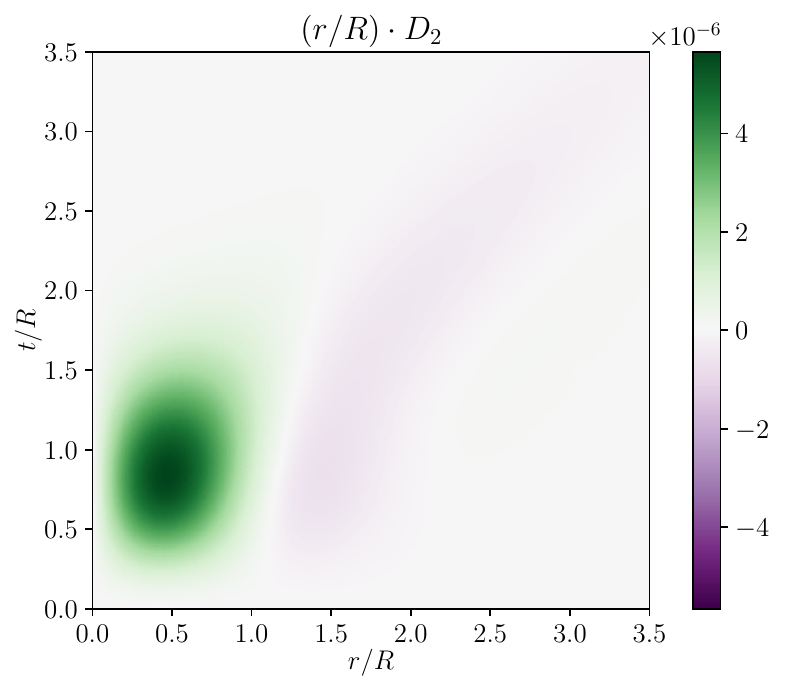}
\includegraphics[width=0.495\textwidth]{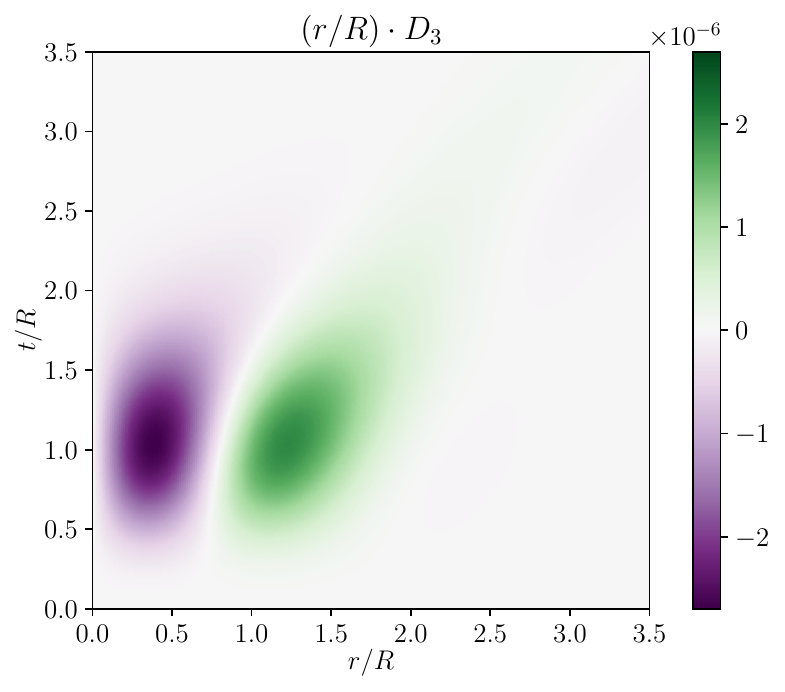}
\includegraphics[width=0.495\textwidth]{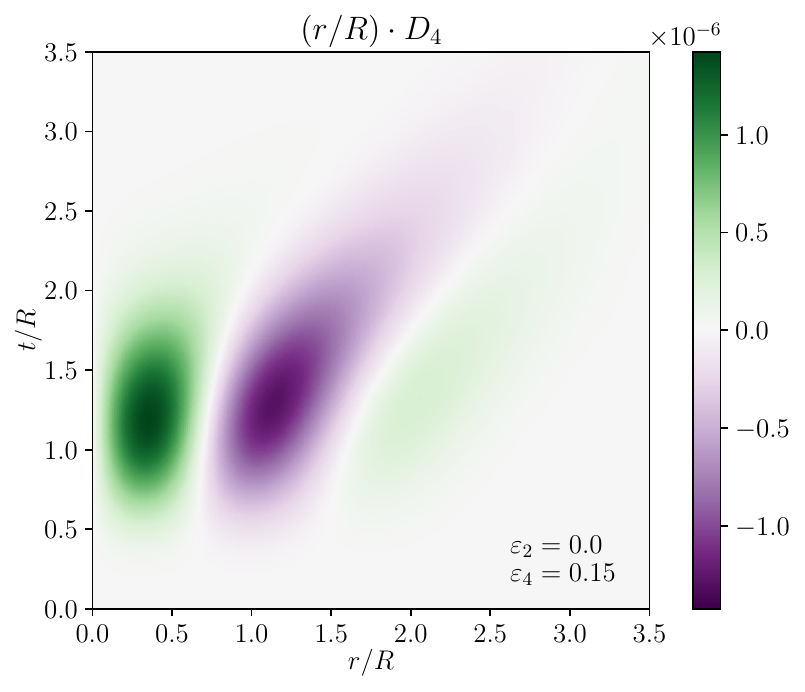}
   \includegraphics[width=0.495\textwidth]{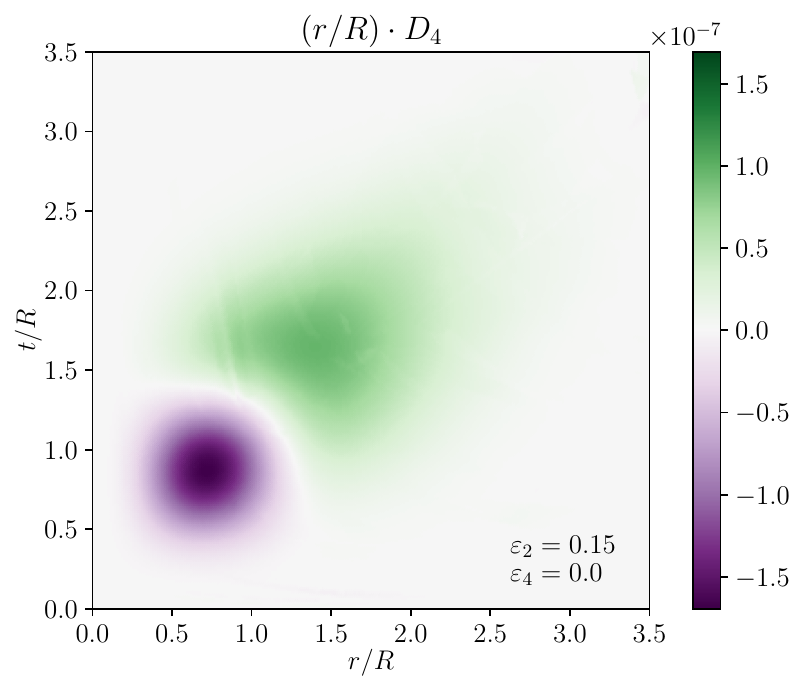}\vspace{-3mm}
\caption{\label{fig:Dn}Angle-averaged local production rates $D_2$ (upper left), $D_3$ (upper right), and $D_4$ for an initial state with $\varepsilon_2=0$ and $\varepsilon_4=0.15$ (bottom left) or with  $\varepsilon_2=0.15$ and $\varepsilon_4=0$ (bottom right).}
\end{figure*}

To probe the temporal and spatial origin of the anisotropic flow buildup better, we study the production rate of each flow harmonic as a function of time and position~\cite{Kurkela:2021ctp,Ambrus:2021fej}. 
This local production rate of $v_n$, averaged over the polar angle of the production point, is quantified by $D_n(t,r)$ introduced in Eq.~\eqref{v_n(t)_O(sigma)}, from which its expression can be read off. 
Figure~\ref{fig:Dn} shows the results of our analytical approach for $D_2$, $D_3$, and $D_4$, for the setups of Sects.~\ref{ss:v2}--\ref{ss:v4} with $\Nresc=0.02$.
Note that we display $D_n(t,r)$ multiplied by $r$, so that the production rate $\partial_t v_n(t)$ of $v_n(t)$ is simply the integral over $r$. 

The three plots (upper row: $D_2$, $D_3$; lower left: $D_4$) showing the linear response of $v_n(t)$ to the corresponding initial $\varepsilon_n$ exhibit similar qualitative features. 
The innermost region of the system --- extending up to $r\simeq R$ in the case of $D_2$, up to $r\simeq 0.7R$ for $D_3$ and $D_4$ --- contributes to $v_n$ with the same sign (positive for $n=2$ and 4, negative for $n=3$) over the whole evolution. 
Further away from the center come regions that contribute with the opposite sign, not much so for $v_2$, more visibly for $v_3$ and $v_4$. 
A third outer region with the same sign as the innermost one is clearly visible in the case $n=4$, and very faintly for $n=3$.
As time passes by, these regions tend to move towards larger $r$ values, but less markedly than in the similar study for energy-weighted flow~\cite{Ambrus:2021fej}.

These space-time dependent $D_n$ underlie the time dependence of the corresponding $v_n(t)$. 
Thus, the change of sign of $v_4(t)$ from positive to negative in the upper panels of Fig.~\ref{fig:v4}, with a derivative that turns negative around $t/R\simeq 1.5$, reflects the progressive dominance of the region at intermediate $r$ in the lower left panel of Fig.~\ref{fig:Dn}. 
Similarly, the (small) decrease of $v_2(t)$ for $t/R\gtrsim 2$ is due to the outer, negatively contributing regions in $D_2(t,r)$. 
In the case $n=3$ the contributions from the various regions exactly cancel out at every $t$ to yield $\partial_t v_3(t) = 0$, while in Ref.~\cite{Kurkela:2021ctp} an ``almost nearly perfect cancellation'' resulting in a very small negative $v_3^E$ value was found.

Eventually, one can also note that the buildup of the linear $D_n$ happens more slowly with increasing $n$, which possibly reflects the scaling behavior $v_n(t)\propto t^{n+1}$ in the few collisions regime~\cite{Borrell:2021cmh}.

The lower right plot of Fig.~\ref{fig:Dn}, showing $D_4$ for the nonlinear response of $v_4$ to an initial $\varepsilon_2$, is completely different, with a clear negative contribution at early times and for $r \lesssim 1.3R$, followed by a positive contribution at later times and for all values of $r$.
In that case one easily checks that the initial eccentricity $\varepsilon_2\neq 0$, irrespective of its sign, i.e.\ the ellipse orientation, generates via the loss term a negative $v_4\propto -\varepsilon_2^2$. 
Simultaneously, the ellipticity decreases in absolute value, due to the $v_2$ which is also created. 
As the negative quadrangular flow $v_4$ develops, it leads to the development of a positive quadrangularity $\varepsilon_4$, which is the seed for the positive contributions to $v_4$ at later times, as seen in the bottom panels of Fig.~\ref{fig:v4}.

\section{Summary}
\label{s:summary}

We have investigated anisotropic flow in the few-rescatterings regime in four models, starting with a transport code with elastic binary scatterings, which serves as the reference including all rescatterings in the system. 
To assess which fraction of the anisotropic flow is carried by particles that escape the system without scattering, we introduced a $2\to 0$ version of the code. 
With the help of a variant of the $2\to 2$ code in which particles that have collided once may no longer rescatter, but are accounted for in the final state, we estimate the amount of anisotropic flow at the ``single-hit'' level.
Eventually, we carried out analytical calculations within Boltzmann kinetic theory, including only the loss term of the binary collision kernel and restricting ourselves to linear order in the cross section. 
Intrinsically the analytical approach and the simulations with the $2\to 0$ kernel are unphysical, since energy and momentum are not conserved in the rescatterings. 
Nevertheless they provide us with a proxy on how much anisotropic flow is created by particles escaping the system without any interaction. 

On the other side, the strength of the analytical calculations is that they yield directly a number of known scaling behaviors of the anisotropic flow coefficients, like their dependence on the initial-state eccentricities or their early-time onset, confirming earlier studies~\cite{Borrell:2021cmh}.
Remarkably, the analytical approach at order ${\cal O}(\sigma)$ yields $v_n=0$ for all odd coefficients, but finite values for even ones, which hints at a fundamental difference between odd and even harmonics. 
Note that we have found elsewhere that odd $v_n$ harmonics can be non-zero at order ${\cal O}(\sigma^2)$~\cite{Borghini:2022qha}. 

For even harmonics ($v_2$, $v_4$, $v_6$), the results of all approaches are very similar when the number of rescatterings in the system is small. 
In the case of $v_4$ and $v_6$, this holds for both the linear flow response $v_n\propto \varepsilon_n$ and the nonlinear response like e.g.\ $v_4\propto\varepsilon_2^2$.
The agreement suggests that in the few-rescatterings regime, the even components of the flow signal are to a large extent carried by particles that flew out of the system without colliding, with an anisotropic escape probability reflecting the asymmetric geometry, as advocated for $v_2$ in AMPT~\cite{He:2015hfa}.

In contrast, for odd harmonics ($v_3$) the results of the $2\to 2$ and $2\to 0$ numerical scenarios differ significantly, even in the very few rescatterings regime. 
Indeed, the former scale roughly linearly with $\Nresc$, while the latter rather scale like $\Nresc^2$. 
That finding in the $2\to 0$ model is consistent with the fact that we find $v_3=0$ in our analytical calculations at order $\sigma$ --- while we found in a parallel study that there is a non-zero $v_3$ at order $\sigma^2$~\cite{Borrell:2021cmh}.
The $v_3$ results from the single-hit model also differ significantly from those of $2\to 2$ simulations.
All in all, the strong dependence of triangular flow on the choice of collision kernel confirms the observation in Ref.~\cite{Kurkela:2021ctp}.
In particular, the discrepancy between the approaches demonstrates that in the case of the odd harmonics, the observed $v_n$ is not driven by the anisotropic-escape probability, but that the fate of particles after they have undergone a collision does matter. 

A clear limitation of the present study is the restriction to a two-dimensional expansion. 
As we explained in the introduction, this is due to the fact that the small $v_3$ values require large statistics, which would be too time-consuming in a three-dimensional study. 
Indeed, we want to emphasize that previous studies~\cite{Romatschke:2018wgi,Kurkela:2018qeb,Kurkela:2019kip,Kurkela:2020wwb,Kurkela:2021ctp,Ambrus:2021fej} of kinetic theory at small opacity relied on solving the (deterministic) Boltzmann equation --- with different collision kernels ---, while here for the first time\footnote{An exception is Ref.~\cite{Borghini:2022qha}, which is restricted to early times.} we used transport simulations at small $\Nresc$. 
This makes it significantly harder to obtain reliably very small $v_n$ values, of order a few $10^{-5}$ at the smallest $\Nresc$ we considered (see Figs.~\ref{fig:v3}, \ref{fig:v4}, \ref{fig:v6}). 
This is even more true in the presence of longitudinal expansion, which dilutes the transverse profile of the system faster, thereby decreasing the anisotropic flow. 

That being told, we may still comment on the results one can anticipate in a three-dimensional expansion, in particular a longitudinally boost-invariant one. 
First, as pointed out in Appendix~\ref{app:v3,v5_no-fact}, the property that $v_3$ and higher odd harmonics vanish in the loss-term-only calculations is sensitive to the presence of a longitudinal direction: 
In a three-dimensional geometry, odd harmonics are probably zero at linear order in $\sigma$ only if the particles are massless and the local momentum distribution in the initial state is independent of position, which is unrealistic. 
That is, we would anticipate that our finding $v_3\propto\Nresc^2$ may not be robust and be replaced by $v_n\propto\Nresc$ for both even and odd $n$. 
It is also clear that rescatterings will generally change the longitudinal components of momenta. 
Thus, it is possible that the agreement we find between all models for even harmonics may not survive the introduction of a third dimension, i.e.\ that the apparent importance of the anisotropic-escape contribution to the coefficients $v_{2n}$ may no longer persist. 
However, we do not see how longitudinal expansion could enhance the effectiveness of the escape mechanism at producing the odd flow harmonics --- although it may decrease the relative importance of the component modeled by the gain term of the Boltzmann equation in some regions of phase space. 

We would thus conclude that the ``escape mechanism'' picture can{\em not\/} account for the whole anisotropic flow signal in systems with very few rescatterings per particle. 
Within our study, the mechanism is efficient for even harmonics, but not for odd ones.  
It also means that the details of the (differential) scattering cross section certainly matter for predicting the value of odd anisotropic flow harmonics, as already hinted at by the results on $v_3$ in Ref.~\cite{Kurkela:2021ctp}, while the even harmonics may be less sensitive. 
To our knowledge, such a difference in the microscopic ``origin'' of even and odd flow harmonics has not been reported before in the framework of transport studies.\footnote{A difference between even and odd harmonics of two-particle azimuthal correlations was found in a study of proton--nucleus collisions within a Color Glass Condensate based approach~\cite{Mace:2018yvl}.} 
Since this will be relevant in systems with small enough multiplicities, our study within a toy transport model (two-dimensional expansion, hard spheres) clearly needs to be replicated with more realistic codes and setups.

\begin{acknowledgments}
We thank Marc Borrell, Kai Gallmeister, Carsten Greiner, Sören Schlichting and Clemens Werthmann for fruitful discussions.
The authors acknowledge support by the Deutsche Forschungsgemeinschaft (DFG, German Research Foundation) through the CRC-TR 211 'Strong-interaction matter under extreme conditions' - project number 315477589 - TRR 211.
Numerical simulations presented in this work were performed at the Paderborn Center for Parallel Computing (PC$^2$) and the Bielefeld GPU Cluster, and we gratefully acknowledge their support.
\end{acknowledgments}

\appendix

\section{Odd flow harmonics in the ``loss term'' scenario}
\label{app:v3,v5}

In this Appendix, we show that the odd flow harmonics $v_3$, $v_5$\dots\ are identically zero when computed at leading order in the cross section with a collision kernel including only the loss term of the Boltzmann equation, Eq.~\eqref{loss-term}. 

\subsection{Factorized initial distribution}
\label{app:v3,v5_fact}

Following Eq.~\eqref{v_n(t)_O(sigma)}, our analytical calculation of $v_n(t)$ involves the integral over the transverse plane
\begin{equation}
\label{appA_eq1}
{\cal I} \equiv \!\int\! f^{(0)\!}({\bf x}-{\bf v}t,{\bf p})_{} f^{(0)\!}({\bf x}-{\bf v}_1t,{\bf p}_1) \,{\rm d}^2{\bf x},
\end{equation}
where the free-streaming distribution has been expressed in terms of the initial condition via Eq.~\eqref{f_f.s.}.
A straightforward change of variable yields
\begin{equation}
\label{appA_eq2}
{\cal I} = \!\int\! f^{(0)\!}({\bf x}-\bm{\xi},{\bf p})_{} f^{(0)\!}({\bf x}+\bm{\xi},{\bf p}_1) \,{\rm d}^2{\bf x}
\end{equation}
with $\bm{\xi}\equiv\frac{1}{2}({\bf v}+{\bf v}_1)t$.

Let us assume right away that the initial-state phase space density factorizes into independent spatial and momentum distributions as in Eq.~\eqref{f^(0)=F.G}.\footnote{The assumption only matters when going from Eq.~\eqref{appA_eq4} to Eq.~\eqref{appA_eq6}.}
In that case the momentum parts are irrelevant for the integral over ${\bf x}$ and the calculation of ${\cal I}$ involves that of 
\begin{equation}
\label{appA_eq3}
{\cal I}' \equiv \!\int\! F({\bf x}-\bm{\xi})_{} F({\bf x}+\bm{\xi}) \,{\rm d}^2{\bf x}.
\end{equation}
Since the integral runs over the whole transverse plane, we may equivalently replace the integrand by its even part:
\begin{equation}
\label{appA_eq4}
{\cal I}' = \frac{1}{2}\!\int\!\big[F({\bf x}\!-\!\bm{\xi})_{} F({\bf x}\!+\!\bm{\xi}) + 
  F(-{\bf x}\!-\!\bm{\xi})_{} F(-{\bf x}\!+\!\bm{\xi}) \big] {\rm d}^2{\bf x}.
\end{equation}
This integral can be further transformed by introducing the even and odd parts of the spatial profile $F$:
\begin{equation}
\label{appA_eq5}
F({\bf x}) = F_+({\bf x}) + F_-({\bf x})
\text{ with }
F_\pm(-{\bf x}) = \pm F_\pm({\bf x}).
\end{equation}
The even ``eccentricities'' of the geometry and its isotropic component are entirely controlled by $F_+$, while $F_-$ accounts for the odd eccentricities.
Replacing $F$ by $F_+ + F_-$ in Eq.~\eqref{appA_eq4}, the integrand yields 8 terms: four of those cancel pairwise and there remains
\begin{equation}
\label{appA_eq6}
{\cal I}' = \!\int\!\big[F_+({\bf x}-\bm{\xi})_{} F_+({\bf x}+\bm{\xi}) + 
F_-({\bf x}-\bm{\xi})_{} F_-({\bf x}+\bm{\xi}) \big] {\rm d}^2{\bf x}.
\end{equation}

As is well established in model studies,\footnote{See also Ref.~\cite{Borrell:2021cmh} for a more formal proof within kinetic theory.} in the absence of initial anisotropic flow a given harmonic $v_n$ can only arise as linear response to a modulation of the initial geometry in the same $n$-th harmonic --- symbolically $v_n\propto \varepsilon_n$ ---, or as quadratic response to two geometrical modulations that combine appropriately --- symbolically $v_n\propto \varepsilon_k\varepsilon_{n-k}$, or more generally (but this case cannot be obtained in the analytical approach of the present paper) $v_n \propto \varepsilon_{k_1}\cdots\varepsilon_{k_m}$ with $k_1+\cdots+k_m = n$.
Since all modulations of even (including 0) resp.\ odd order are accounted for by $F_+$ resp.\ $F_-$, one sees that the products $F_+F_+$ or $F_-F_-$ in the integrand of Eq.~\eqref{appA_eq6} can yield the necessary contributions to $v_n$ of the kind $\varepsilon_n$ or $\varepsilon_k\varepsilon_{n-k}$ for any even harmonic $n$, but not for odd $n$. 
That is, the integral ${\cal I}'$ does not depend on the momentum azimuths $\varphi_{\bf p}$, $\varphi_1$ (on which $\bm{\xi}$ implicitly depends) in such manner that after multiplying with the M{\o}ller velocity and $\cos(n\varphi_{\bf p})$ and integrating over these azimuths, there could result a non-zero $v_n$ when $n$ is odd.

To conclude, note that our proof does not explicitly use the dimensionality of the system, nor does it make any assumption on the particle mass. 
However, it assumes that the collision kernel does not include quantum effects.

\subsection{Position-dependent initial momentum distribution}
\label{app:v3,v5_no-fact}

Coming back to a two-dimensional setup with massless particles, let us drop the factorization assumption~\eqref{f^(0)=F.G} for the initial phase space distribution. 
That is, we now write the initial distribution as
\begin{equation}
\label{f^(0)=F.G_bis}
f^{(0)}({\bf x},{\bf p}) = F({\bf x})_{}G({\bf p};\Lambda({\bf x})),
\end{equation}
where $\Lambda({\bf x})$ symbolizes the dependence of the local momentum distribution on the position in the transverse plane: this could for instance be a local saturation scale or a local temperature. 
As previously, we may still assume without loss of generality that $G$ is normalized to unity at every position ${\bf x}$ when integrating over the whole momentum space --- as is e.g.\ the case if it is a thermal Boltzmann distribution $G({\bf p};\Lambda({\bf x}))\propto {\rm e}^{-|{\bf p}|/\Lambda({\bf x})} / \Lambda({\bf x})^2$.
The crux is that since $G$ is assumed to be isotropic in momentum space, i.e.\ only depends on the modulus $|{\bf p}|$ the normalization of $G$ translates at once into 
\begin{equation}
\label{int[G(p)]}
\int\!G({\bf p};\Lambda({\bf x}))\,|{\bf p}|\,{\rm d}|{\bf p}| = \frac{1}{2\pi},
\end{equation}
which holds irrespective of whether or not the local momentum distribution is position-dependent. 

Let us go back to Eq.~\eqref{v_n(t)_O(sigma)} which gives $v_n(t)$ at order ${\cal O}(\sigma)$ in the loss-term approach. 
The free-streaming distributions $f_{\rm f.s.}$ in the integrand are evaluated at positions ${\bf x}-{\bf v}t'$ resp.\ ${\bf x}-{\bf v}_1t'$ [cf.\ Eq.~\eqref{f_f.s.}] that only involve the azimuths of the momenta ${\bf p}$, ${\bf p}_1$. 
The integrals over the moduli $|{\bf p}|$ and $|{\bf p}_1|$ can thus be performed at once using Eq.~\eqref{int[G(p)]}. 
That is, effectively the precise form of $G$ --- especially its dependence or not on position --- does not matter for $v_n(t)$ in our analytical approach. 
Thus, if the odd $v_n(t)$ harmonics vanish for a position-independent initial momentum distribution, as shown in Appendix~\ref{app:v3,v5_fact}, then this remains true if $G$ depends on ${\bf x}$. 

Note that the above proof does not readily generalize to a three-dimensional system nor to massive particles: 
in such cases, the M{\o}ller velocity in the integrand of Eq.~\eqref{v_n(t)_O(sigma)} takes a more complicated form, and in particular it depends on $|{\bf p}|$ and $|{\bf p}_1|$, so that Eq.~\eqref{int[G(p)]} can no longer be used. 
Similarly, it does not hold either for energy-weighted flow coefficients (like $v_3^E$), because in that case an extra factor of $|{\bf p}|$ enters the integrand on the right-hand side of Eq.~\eqref{v_n(t)_O(sigma)}, which again prevents the use of Eq.~\eqref{int[G(p)]}.
This is consistent with the findings in Ref.~\cite{Borghini:2022qha} in which $v_3$ is first non-zero at order ${\cal O}(\sigma^2)$, while $v_3^E$ is already finite at order ${\cal O}(\sigma)$.

\section{Scaling of $v_2$ with the inverse Knudsen number}
\label{app:v2_vs_sigma}

\begin{figure}[t]
	\includegraphics[width=\linewidth]{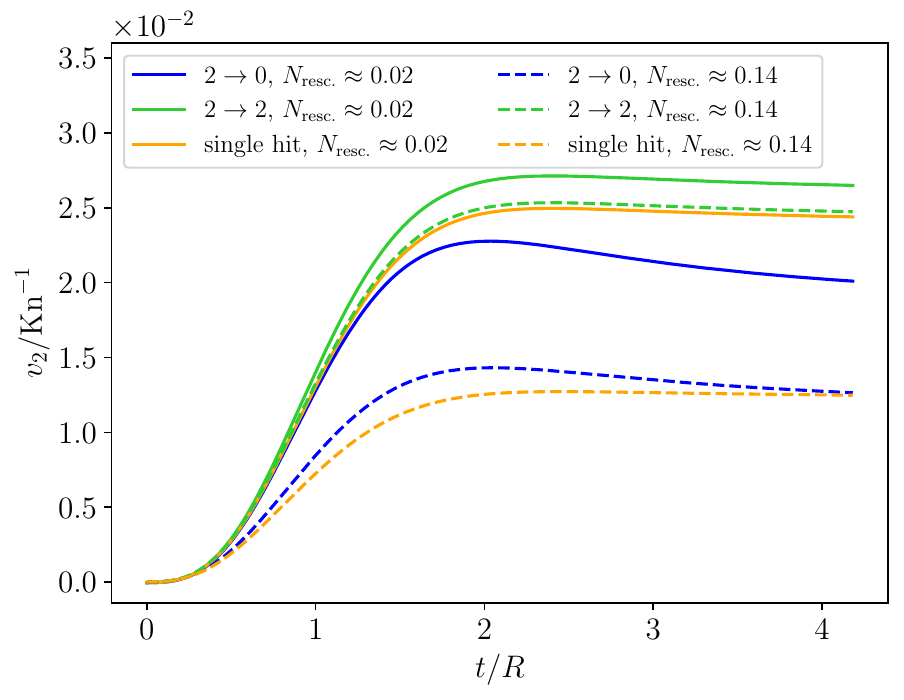}\vspace{-3mm}
	\caption{\label{fig:v2/sigma}Ratio $v_2(t)/{\rm Kn}^{-1}$ in systems with initially $\varepsilon_2=0.15$ and on average $\Nresc\approx 0.02$ (full lines) or $0.14$ (dashed) rescatterings per particle for the three scenarios of the transport cascade: 
		$2\to 2$ (green), $2\to 0$ (blue), single hit (orange).}
\end{figure}

In Fig.~\ref{fig:v2/sigma} we show $v_2(t)$ divided by the inverse Knudsen number for the three scenarios of our transport code and for the two values $\Nresc\approx 0.02$ and $0.14$. 
While ${\rm Kn}^{-1}$ is approximately the same in all three models for $\Nresc\approx 0.02$, it differs significantly between the $2\to 2$ model and the other two at $\Nresc\approx 0.14$. 

As could be expected from Fig.~\ref{fig:v2/Nresc}, one finds that $v_2\propto{\rm Kn}^{-1}$ to better than 10\% accuracy in the full $2\to 2$ simulations, consistent with the equally good scaling with $\Nresc$ and the fact that $\Nresc$ and ${\rm Kn}^{-1}$ are proportional. 
On the other hand, it is clear that the scaling is not so good for the $2\to 0$ scenario --- and even less in the single-hit model.

\section{Results for $\Nresc \approx 0.35$}
\label{app:Nresc=0.35}

\begin{figure}[tb!]
\includegraphics[width=\linewidth]{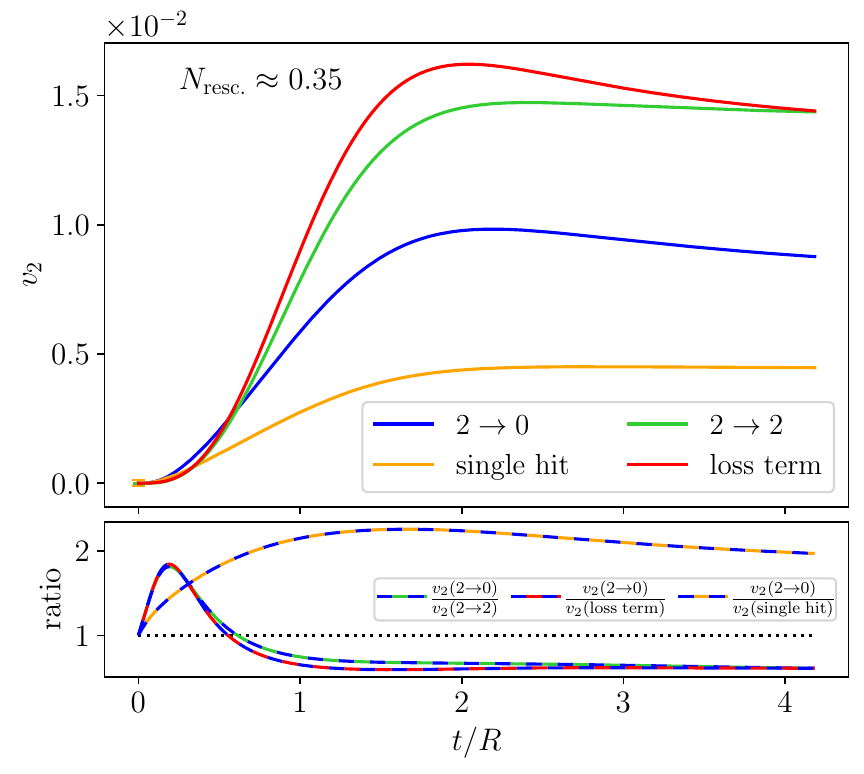}\vspace{-3mm}
\caption{Time dependence of elliptic flow $v_2$ in systems with initially $\varepsilon_2=0.15$ and on average $\Nresc\approx 0.35$ rescatterings per particle, in systems 
The green curves are for systems with elastic binary scatterings, the blue lines for the $2\to 0$ scenario, the orange ones for the single-hit model, and the red lines show the analytical result~\eqref{eq:v2(t)}.}
\label{fig:v2bis}
\end{figure}

\begin{figure*}[tb!]
\includegraphics[width=0.495\linewidth]{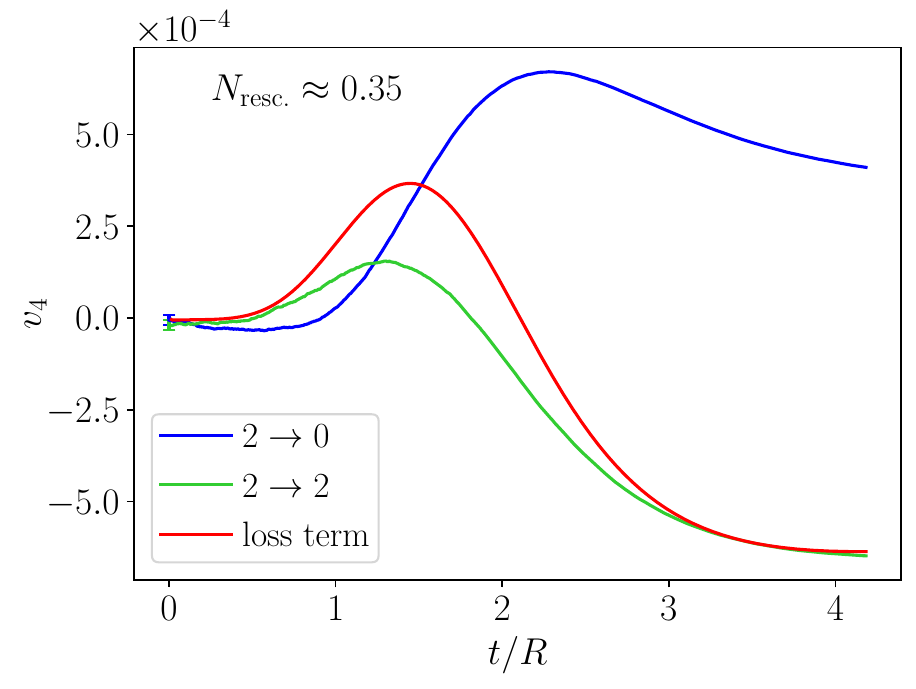}
\includegraphics[width=0.495\linewidth]{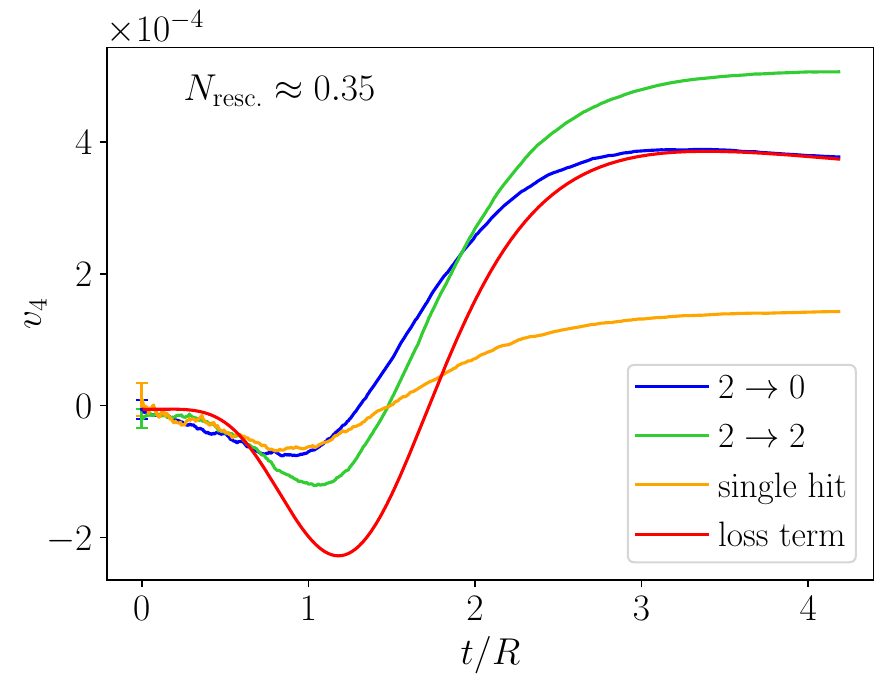}\vspace{-3mm}
\caption{Time-dependence of quadrangular flow $v_4$ in systems with $\Nresc\approx 0.35$ rescatterings per particle on average. 
Left: system with an initial $\varepsilon_4 = 0.15$ and $\varepsilon_2 = 0$; right: system with an initial state $\varepsilon_2 = 0.15$ and $\varepsilon_4 = 0$. 
The green curves are for systems with elastic binary scatterings, the blue lines for the $2\to 0$ scenario, the orange one for the single-hit model, and the red lines show the analytical results~\eqref{eq:v4_e4(t)} (left) or Eq.~\eqref{eq:v4_e2(t)} (right).}
\label{fig:v4bis}
\end{figure*}

In this Appendix we provide for the sake of reference results for $v_2$ (Fig.~\ref{fig:v2bis}) and $v_4$ (Fig.~\ref{fig:v4bis}) for systems in which the mean number of rescatterings per particle is about 0.35. 
For the $2\to 0$ resp.\ single-hit scenarios, this means that approximately $70\%$ of the particles disappear resp.\ become transparent over the system evolution. 
Accordingly, the assumption underlying the analytical calculations, that the phase-space distribution $f(t,{\bf x},{\bf p})$ deviates negligibly at all times from the free-streaming distribution $f_{\rm f.s.}(t,{\bf x},{\bf p})$ with the same initial condition, is clearly non fulfilled. 
In addition, it is somewhat clear that if 70\% of the particles scatter once in the $2\to 0$ or single-hit models, then a significant fraction of them would actually collide several times in the $2\to 2$ model: extrapolating the straight-line fit in Fig.~\ref{fig:Nresc_vs_Kn} indeed gives $\Nresc\approx 1.67$ for a $2\to 2$ system with the same initial input Knudsen number as used in the $2\to 0$ or single-hit simulations. 
That is, it is clear from the start that the ``full'' $2\to 2$ and ``truncated'' $2\to 0$ or single-hit systems that lead to $\Nresc\approx 0.35$ are extremely different. 

The two plots displaying ``linear'' flow response, namely $v_n$ for an initial non-zero $\varepsilon_n$ with $n=2$ (Fig.~\ref{fig:v2bis}) or $n=4$ (Fig.~\ref{fig:v4bis} left), are similar: 
The results from the simulations with the $2\to 2$ collision kernel (green lines) and the $2\to 0$ scenario (blue lines) largely differ, by roughly $40\%$ in the case of $v_2$, and even yielding signals with opposite signs in the case of $v_4$. 
In contrast, the analytical results are remarkably close to those from the transport calculations with $2\to 2$ scatterings, in particular the final values of $v_2$ or $v_4$, which in our view should probably not be over-interpreted.
As mentioned in Sect.~\ref{ss:v4}, the agreement for $v_4$ disappears at higher $\Nresc$ values, since the $2\to 2$ results become positive. 

Going to the right panel of Fig.~\ref{fig:v4bis} showing the nonlinear response $v_4\propto \varepsilon_2^2$, we just note that the agreement between the three approaches is quite good, again without attempting to interpret it.

\section{Alternative distribution function}
\label{app:different_dist}

\begin{figure*}[tb!]
\includegraphics[width=0.495\linewidth]{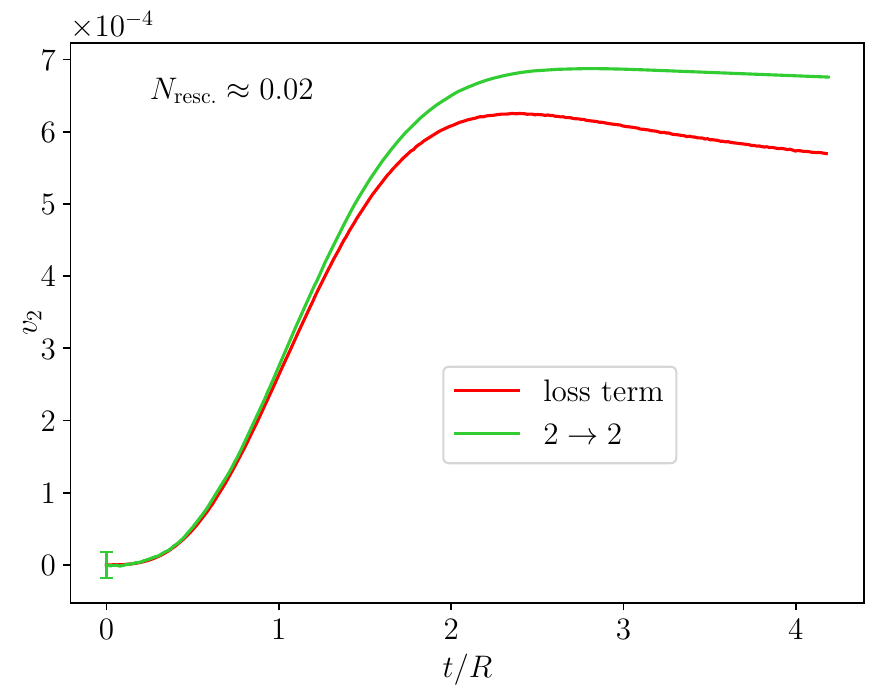}
\includegraphics[width=0.495\linewidth]{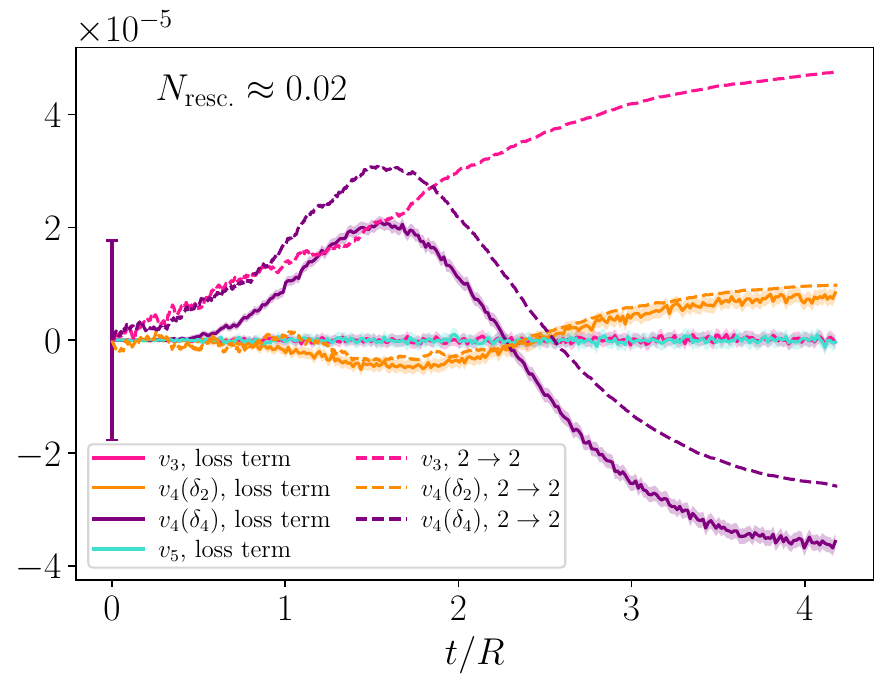}\vspace{-3mm}
\caption{\label{fig:vn_2010Dist}Time dependence of anisotropic flow harmonics for the initial geometric profile~\eqref{eq:different_distribution_function} in systems with $\Nresc \approx 0.02$.
Left:  $v_2$ for $\delta_2=0.15$ (green: numerical simulations, red: semi-analytical approach). 
Right: $v_3$ for $\delta_3=0.15$ (magenta), $v_4$ (orange: for $\delta_2=0.15$, purple: for $\delta_4=0.15$), and $v_5$ (cyan, for $\delta_5=0.15$) in numerical simulations (dashed lines) or computed with Vegas (full lines; the color bands show the $3\sigma$ error of the MC integration).}
\end{figure*}

To check whether our main results are specific to our choice of initial profile~\eqref{eq:distribution_function}, we repeated the calculations in the few-rescatterings regime $\Nresc \approx 0.02$ for a slightly different geometry, namely 
\begin{equation}
	\tilde{F}(r,\theta) =  \frac{N \sqrt{1-\delta_j^2}}{2\pi R^2} 
	\exp\bigg(\!\!-\!\frac{r^2 [1+\delta_j \cos(j\theta)]}{2 R^2}\bigg),
	\label{eq:different_distribution_function}
\end{equation}
which was used in the fluid-dynamical regime in Ref.~\cite{Alver:2010dn}.
A drawback of this distribution is that a given $\delta_j$ contributes to several eccentricities $\varepsilon_n$, namely for all $n$ that are multiples of $j$. 
On the other hand, the density~\eqref{eq:different_distribution_function} is positive definite irrespective of the parameter values. 

As we could not perform all integrals with the distribution function~\eqref{eq:different_distribution_function} analytically, we used the Vegas Monte Carlo (MC) integration method~\cite{Vegas3.5.3} to evaluate the flow coefficients.
In addition, we performed transport simulations only with the $2\to 2$ collision kernel.

Figure~\ref{fig:vn_2010Dist} shows our results for the flow coefficients $v_n$ with $n\in\{2,3,4,5\}$.
The odd harmonics ($v_3$, $v_5$) from the semi-analytical approach are zero within the error bars of the MC integration, as expected from Appendix~\ref{app:v3,v5}, while the triangular flow $v_3$ is clearly non-zero in the transport simulations. 
In contrast to this mismatch for the odd harmonics, the $v_2$ and $v_4$ signals in a system with an initial $\delta_2$ are in nice agreement in the two approaches. 
The agreement is less good for the $v_4$ from an initial $\delta_4$, but this may be due to the numerical noise in the simulations at early times, since at later times the two curves run parallel to each other. 
In addition, the overall shapes of $v_2$ and $v_4$ (either from a non-zero $\delta_2$ or a non-zero $\delta_4$) are similar to those found in Sects.~\ref{ss:v2} and \ref{ss:v4} with the distribution~\eqref{eq:distribution_function}.

\end{document}